\documentclass[11pt]{article}

\usepackage[
    left=1.5in,
    right=1.5in,
    top=1in,
    bottom=1in
]{geometry}
\usepackage{setspace}
\usepackage{microtype}

\usepackage[T1]{fontenc}
\usepackage{textcomp}
\usepackage{amsmath,amssymb}
\usepackage[utf8]{inputenc} % allow utf-8 input
\usepackage[T1]{fontenc}    % use 8-bit T1 fonts
\usepackage{hyperref}       % hyperlinks
\usepackage{url}            % simple URL typesetting
\usepackage{booktabs}       % professional-quality tables
\usepackage{amsfonts}       % blackboard math symbols
\usepackage{nicefrac}       % compact symbols for 1/2, etc.
\usepackage{microtype}      % microtypography
\usepackage[table]{xcolor}         % colors
\usepackage{graphicx}
\usepackage[table]{xcolor}
\usepackage{colortbl}
\usepackage{array}
\usepackage{amsmath}
\usepackage{amsfonts}
\usepackage{tabularx}
\usepackage{makecell}
\usepackage{pifont}
\usepackage{threeparttable}
\usepackage[numbers,sort]{natbib}
\usepackage{float}

\usepackage{graphicx}
\usepackage{booktabs}
\usepackage{afterpage}

\usepackage[
  aboveskip=6pt,
  labelfont=bf,
  labelsep=period,
  justification=raggedright,
  singlelinecheck=false
]{caption}

\usepackage[
  format=hang,
  singlelinecheck=false,
  font={small},
  labelfont=bf
]{subfig}

\usepackage[ruled,vlined]{algorithm2e}

\usepackage[dvipsnames]{xcolor}

\usepackage{url}
\usepackage{hyperref}
\usepackage{cleveref}

\Crefname{equation}{Eq.}{Eqs.}
\Crefname{figure}{Fig.}{Figs.}
\Crefname{table}{Table}{Tables}
\Crefname{section}{Sec.}{Secs.}

\newcommand{\dd}{\mathrm{d}}

\graphicspath{{finalfigures/}}

\title{From Local Learning to Global Prediction Through Layered Surprise Cascades}

\author{%
Andrew L.~Smith\textsuperscript{0,3},
Linxing Preston Jiang\textsuperscript{1,2},
Jason K.~Eshraghian\textsuperscript{0},\\
Matthew S.~Bull\textsuperscript{1,3,\ensuremath{\dagger}},
\& Stefano Recanatesi\textsuperscript{3,4,\ensuremath{\dagger}}
\\[1em]
\small
\textsuperscript{0}Electrical Engineering Department,
University of California Santa Cruz\\
\textsuperscript{1}Computational Neuroscience Center,
University of Washington\\
\textsuperscript{2}Computer Science Department,
University of Washington\\
\textsuperscript{3}Allen Institute for Neural Dynamics\\
\textsuperscript{4}Technion Israel Institute of Technology\\[0.4em]
\textsuperscript{\ensuremath{\dagger}} Equal senior contribution
}

\date{May 2025}

\begin{document}

\maketitle
% \linenumbers

\begin{abstract}
Hierarchical predictive coding proposes a compelling hypothesis of brain computation, suggesting that the cortex builds layered predictions to minimize surprise. Yet most models rely on error-coding neurons or generative modeling of unclear biological plausibility. Here, we examine a biologically plausible framework in which the functional goals of predictive coding emerge from local contrastive learning and simple activity cancellation. Building on recent machine learning advances, we present a recurrent variant of the Forward-Forward (FF) algorithm with an inverted objective that increases activity for negative data. This setup yields predictive representations across layers, capturing hallmark features of cortical computation such as top-down modulation and surprise signaling. Our results suggest that key principles of predictive coding can emerge from simple, local learning rules, offering a new bridge between neuroscience and machine learning.
\end{abstract}

\paragraph{Author summary}
A growing body of evidence has shown that neural dynamics suppress expected inputs and amplify unexpected ones, creating a layered cascade of surprise across the cortex. This is especially clear in the visual system, where early areas suppress predictable features, and deeper areas respond more strongly to surprising events. Although many models can reproduce this behavior, most rely on biologically implausible mechanisms such as error-detecting neurons or mirrored feedback pathways.

We investigated whether such surprise cascades could emerge from simpler, local learning rules already used in contrastive learning, focusing on the Forward-Forward algorithm. 

Surprisingly, despite label information being delivered from the top, the model learns to cancel predictable activity from the bottom up. The result is a bottom-up cascade of cancellation and surprise that closely mirrors dynamics observed in the visual cortex.

Our model requires only local synaptic learning and a simple global signal, making its neural implementation biologically plausible. Crucially, we demonstrate that our local contrastive objective is mathematically equivalent to a three-factor Hebbian learning rule, where synaptic updates are determined by pre-synaptic activity, post-synaptic activity, and a global gating signal. This grounds our model in established principles of synaptic plasticity. It offers a concrete link between local contrastive learning and hierarchical brain dynamics, with clear experimental predictions.

\section{Introduction}

The brain continuously predicts sensory inputs, allowing organisms to anticipate events and respond efficiently to their environment \cite{wolpert_internal_1998,kawato_internal_1999,schenck_adaptive_2008}. Predictive processing theories suggest that the neocortex constructs hierarchical internal models that encode statistical regularities of the world, using prior experiences to generate expectations about future stimuli \cite{Jiang2022,rao_predictive_2022,millidge_predictive_2022, bassett_network_2017}. When incoming sensory signals align with these expectations, neural activity is suppressed, whereas deviations elicit enhanced responses, reflecting surprise or prediction errors \cite{Gilbert_Li_2013,Jordan_Keller_2020}.

Across the mammalian visual cortex, recordings reveal an ascending cascade of neural activity: primary areas suppress predictable stimulus features, and this suppression – together with enhanced responses to unexpected features – appears one stage later in secondary and tertiary areas \cite{chaudhuri_large-scale_2015,khan_contextual_2018,froudarakis_visual_2019,piet_behavioral_2023}. Any mechanistic account of predictive processing must therefore explain how such bottom-up waves of suppression-and-surprise can emerge from the local circuitry, when it is thought that modulating signals are passed top down \cite{mechelli_where_2004,badre_frontal_2018}.

Classical predictive coding models propose that top-down pathways generate explicit predictions, while bottom-up signals convey prediction errors to update internal models \cite{rao_predictive_1999}. However, cortical implementations rely on dedicated error neurons, symmetric feed-forward and feedback weights, and iterative message passing – circuit motifs that remain unverified in cortex \cite{lillicrap_random_2016}.

Recent advances in machine learning have suggested new contrastive forms of learning that have proven powerful in a number of tasks (image classification, sequence learning, generative modeling, etc.) \cite{oord2019representationlearningcontrastivepredictive, chen2020simpleframeworkcontrastivelearning, he2020momentumcontrastunsupervisedvisual, grill2020bootstraplatentnewapproach}. Some hierarchical contrastive networks dispense with explicit error neurons and weight symmetry, and have shown promise for unsupervised representation learning \cite{oord2019representationlearningcontrastivepredictive, hinton_forward-forward_2022, ororbia_predictive_2023, ororbia_learning_2023}. Yet, despite their appeal for modeling latent sensory structure, it remains unknown whether the activity they produce mirrors the ascending pattern of cancellation and surprise observed across cortical areas.

This raises a central question: can existing locally contrastive models, operating without error neurons or weight symmetry, generate the ascending cascade of suppression and surprisal observed across cortical areas? Any satisfactory answer must ground the rule in local three-factor Hebbian plasticity and reproduce the cascade as an emergent circuit dynamic.

In this work, we demonstrate that predictive representations and surprise responses can emerge naturally from local learning rules when contrastive learning is applied hierarchically. Rather than relying on explicit error neurons or global top-down predictions, we show that a stacked contrastive learning framework—where each layer independently minimizes activity for predictable inputs—produces neural dynamics consistent with hierarchical predictive coding. 

When implemented in a network architecture with layered connectivity resembling the visual cortex, this framework gives rise to spatiotemporal prediction as an emergent property, with neurons encoding a joint representation of both space and time. Crucially, these effects arise purely from spatially and temporally local contrastive learning, without requiring explicit error signals, top-down feedback, or symmetric weight transport.

Building on the Forward-Forward algorithm, we introduce a recurrent variant with an inverted objective—the Inverted Forward-Forward (IFF) model—designed specifically to incentivize activity cancellation for expected stimuli. We tested this framework using a bidirectionally connected 5-layer neural network trained with local contrastive learning, where each layer independently adjusts its activity to minimize predictable inputs and enhance responses to surprising ones. We show that this architecture naturally produces predictive suppression and cancellation signals, where expected stimuli elicit reduced neural activity while unexpected inputs drive amplified responses.

 These emergent signals closely resemble hierarchical surprise dynamics observed in cortical circuits \cite{siegle_survey_2021, garrett_stimulus_2023, piet_behavioral_2023}. By analyzing network activations over time, we demonstrate that information flow in the model follows a structured cancellation pattern, with surprise responses propagating through layers in a manner consistent with both predictive processing theories and neural dynamics in the mouse visual cortex. Additionally, we establish a theoretical link between our contrastive learning rule and three-factor Hebbian plasticity, reinforcing its biological plausibility \cite{halvagal_combination_2023}. Together, these results suggest that hierarchical contrastive learning provides a viable alternative to classical predictive coding, enabling predictive computations to emerge without explicit error neurons, top-down predictions, or symmetric weight transport.

\section{Results}
We begin by detailing the model architecture and implementation, including the learning rule and training protocol. We then validate the emergence of a hierarchical cascade of cancellation and surprise signals, analyze the underlying network dynamics, and finally examine the theoretical properties of the learning rule. The overarching focus is to demonstrate that simple local learning mechanisms give rise to layered predictive computations via hierarchical cancellation mechanisms.

\subsection{Model Design}

\begin{figure}[h]
\centering
\includegraphics[width=\linewidth]{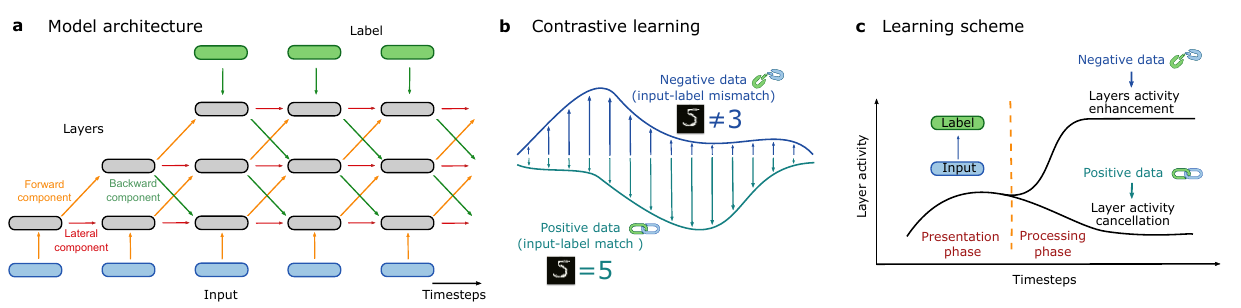}
\caption{{\footnotesize Simple illustrations representing model architecture and learning scheme: (a) Model architecture is shown where data inputs are clamped to the bottom and label inputs are clamped to the top of the network. (b) Forward-Forward contrastive learning schematic with definition of positive and negative datasets, where the label mismatches or matches the sensory input. The y-axis is the energy of a model. The x-axis is the internal representation space. For negative data, activations are raised in arbitrary places in the representation space. For positive data, activations are lowered in arbitrary places. (c) Learning scheme of the model is shown where the training phase proceeds in two steps - the presentation and processing phase respectively. In the processing phase, positive data should have a low activity, whereas negative data should have a high activity.}}
\label{fig:1}
\end{figure}

Our approach builds on the Forward-Forward (FF) model \cite{hinton_forward-forward_2022, ororbia_predictive_2023, ororbia_learning_2023}, a backpropagation-free learning paradigm categorized as a form of contrastive learning. The model architecture is a hierarchical network composed of multiple layers, with label information clamped at the top-most layer and sensory input clamped at the bottom layer (\cref{fig:1}a). In this framework, the label acts as a second input, and the network's output is defined as the layer-wise magnitude of neuronal activity.

This activity magnitude serves as a proxy for compatibility between the label (top input) and the data (bottom input). Higher activity magnitudes suggest a mismatch or "surprise," while lower magnitudes imply consistency between the two inputs.

The network evolves dynamically in the time domain. At each timestep, the activity of a given layer is updated based on the activity of adjacent layers from the previous timestep. Specifically, each neuron receives pre-synaptic inputs from three sources: the layer below, the layer above, and from itself (i.e., recurrent connections). For the bottom and top layers—where one of the adjacent layers is absent—presynaptic input is substituted with the data input (bottom) or the label input (top), respectively.

Training relies on contrastive learning using two types of datasets: positive and negative. A positive sample consists of a correctly paired input and label, while a negative sample consists of a mismatched pair. During training, the model increases layer activity for negative samples, interpreted as an increase in "surprise" (\cref{fig:1}b). Conversely, for positive samples, activity is reduced, indicating diminished surprise and better alignment between label and input.

In this architecture, each layer operates as an independent learning unit, integrating three types of inputs—bottom-up (from the layer below), top-down (from the layer above), and lateral (from its own prior state)—to compute a level of activity. This activity reflects the degree of alignment between the data input and the label input, even if the layer is not directly adjacent to either.

The learning objective is defined at the level of individual layers, based on their activation vectors at time $t'$, denoted $\vec{x}{\textrm{layer}}(t')$:
\begin{equation}
\mathcal{L}_{\textrm{layer}}
=
\sigma\left(
(-1)^{\eta}
\left[
\vec{x}_{\textrm{layer}}^T(t')
\vec{x}_{\textrm{layer}}(t')
-\theta
\right]
\right)
\label{eq:1}
\end{equation}

Here, $\eta=0$ for positive samples (matching input and label) and
$\eta=1$ for negative samples (mismatched input and label).
The parameter $\theta\in\mathbb{R}$ defines the activity margin, and
$\sigma(z)=\log(1+e^z)$ is the softplus function.

Thus, positive samples incur the loss
$\sigma(\|\vec{x}_{\textrm{layer}}\|_2^2-\theta)$,
whereas negative samples incur the loss
$\sigma(\theta-\|\vec{x}_{\textrm{layer}}\|_2^2)$.
Minimizing these terms respectively drives positive activity below the
threshold and negative activity above the threshold. The function $\sigma$ denotes a soft-plus non-linearity, which ensures differentiability and bounded curvature for stable learning.

Although global supervisory signals are often considered biologically implausible, the role of $\eta$ here is minimal—it serves as a binary global indicator that modulates learning uniformly across the network. This design is consistent with biological principles, where neuromodulatory systems (e.g., dopaminergic or serotonergic signals) can diffusely influence activity across large neural populations through volume transmission. Thus, the inclusion of $\eta$ maintains a level of biological plausibility (see Section~\ref{bioplaus}).

\subsection{Model Implementation}
The training protocol consists of performing simultaneous forward passes for both positive and negative examples. Crucially, the model operates solely through forward dynamics at all stages—there are no backward weight updates or error propagation steps. This constraint aligns the training mechanism with biological plausibility, as it avoids the requirement for symmetric weight transport or non-local information. Learning occurs only via adjustments to local synaptic weights associated with top-down, bottom-up, and lateral connections.

Furthermore, we demonstrate that the learning rule underlying this architecture is mathematically equivalent to a class of Hebbian learning rules under certain conditions (see Appendix \cref{app:A}). The network learns to integrate bottom-up input with top-down label signals to produce activity patterns that differentiate between matched and mismatched input-label pairs.

To enhance similarity to biological signal processing, we introduce a revised training schedule (\cref{fig:1}c). Training is divided into two temporal phases: a presentation phase where only the input is provided to the network, followed by a processing phase during which the label is introduced. Each phase comprises a fixed number of timesteps (10 and 15 timesteps, respectively). No weight updates occur during the presentation phase; learning is triggered only in the processing phase, once label information is available.

During training, each layer’s activity increases or decreases depending on whether the label matches the previously presented input. This mirrors cortical computations in which mismatches between sensory input and internal predictions (reflected here by the label) produce elevated activity or "surprise." Conversely, matched signals reduce activity, indicating concordance between bottom-up and top-down information.

Inference follows the same two-phase dynamic. The input is first presented, followed by the label, but all synaptic weights remain fixed. Class identity is then inferred from the latent activity patterns observed during the processing phase.

\begin{figure}[h]
\centering
\includegraphics[width=\linewidth]{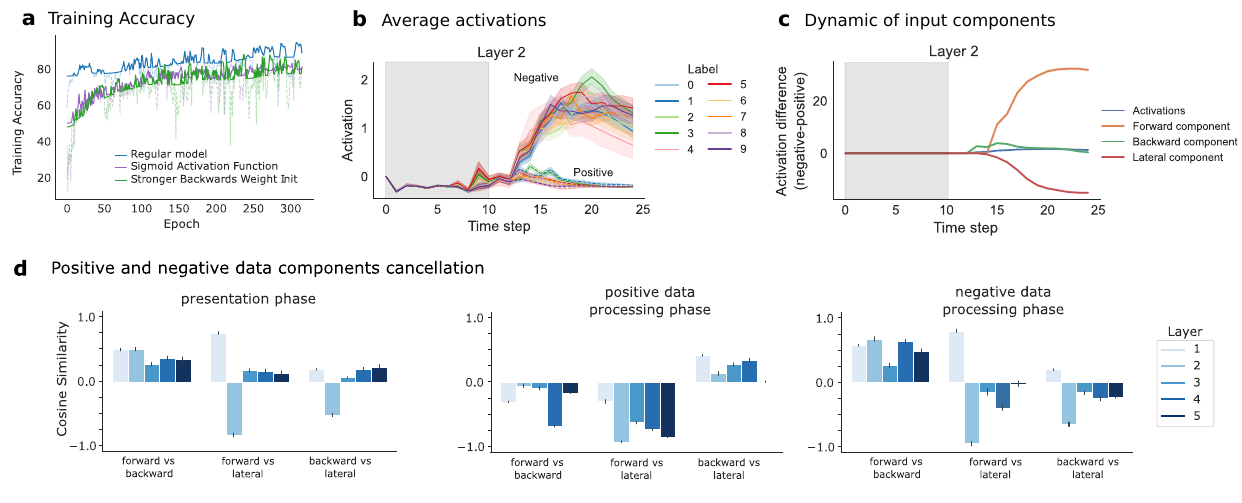}
\caption{{\footnotesize Figures for validation accuracy, layer-wise activation progression throughout time, layer-specific cancellation, and cancellation patterns between the components forming a layer's activation update: (a) Accuracy (y-axis) over time (x-axis) is shown for various configurations of the network. We show a deterioration for both a stronger label clamped weight initialization, as well as with a sigmoid activation instead of leaky ReLU. (b) For the second layer in the network, the average activations (y-axis) obtained over 1000 samples are shown per class over time (x-axis), for both positive and negative data. Negative data induces a large and sustained surprise signal with rising activities after the label is shown. Positive data has a very small surprise and returns to baseline low activity. (c) For the second layer in the network, the activations for positive data averaged across 1000 samples, are broken into their pre-synaptic components (y-axis) and plotted across time (x-axis), which show strong cancellation as indicated by the resultant summed post-synaptic activity (blue). (d) Cosine similarity (y-axis) of positive and negative data activations for the presentation and processing phase (before and after label presentation). The alignment patterns across components highlight the different cancellation profile at work during the processing of positive versus negative data. For positive data, forward opposes lateral and backward synaptic connections in all but layer 2. Layer 2 exhibits more constant behavior independent of the data regime (positive or negative), demonstrating strong forward vs lateral anti-alignment in all cases.}}
\label{fig:2}
\end{figure}

We train the model on the MNIST dataset following the scheme highlighted in \cref{fig:1}. For every iteration, a single MNIST image is selected and presented as an input to the network (presentation phase of 10 timesteps). Following this presentation phase, the label is introduced, while still presenting the image, and the network processes both input and label information (processing phase of 15 timesteps). We first focus on the spatial integration of bottom-up and top-down information flows. Learning follows as per \Cref{eq:1} and accuracy is computed as outlined in \cite{hinton_forward-forward_2022}: for each input image $x$ all possible labels (classes 0 to 9) are introduced to the network, we deem the input image to be accurately processed if the surprise for the correct label is lower than for any other label.

We train a 5 layer network with 700 neurons per layer minimizing \Cref{eq:1}. We used RMSProp as the optimizer with learning rate $5\cdot 10^{-5}$, batch size 500, and Leaky ReLU as the transfer function for all units. We use no momentum and applied a stopgrad operation to all adjacent layer activations to prevent the parameter gradients from growing beyond one-step. Weight initializations and further details can be found in the available repository \footnote{https://github.com/and-rewsmith/RecurrentForwardForward}. The 5-layer, 25 timestep model achieved 95\% test accuracy upon training (\cref{fig:2}a). Different activation functions were attempted, and sigmoid consistently showed to perform worse than ReLU derivatives (\cref{fig:2}a).

\subsection{Hierarchical emergence of surprise and cancellation signal}
\label{sec:hier}
By analyzing layer activity via L2 norm over time, we were able to confirm that the model learned to dynamically suppress neural activity across both layers and time whenever the input image matched the respective label (\cref{fig:2}b). The difference between negative and positive activations showed a clear divergence upon label presentation (\cref{fig:2}b). This trend was the result of the contribution of multiple input components to the layers (\cref{fig:2}c). Notably, the forward component representing the input from lower layers was the only significantly stronger component for negative versus positive data, suggesting a leading role of this component in driving the increased activity for surprise signals. This was true across all layers (\cref{fig:SI_2.0}).

In order to understand how these input components were driving the increase in activity for negative data (surprise signal) -- and decrease in activity for positive data (the cancellation upon label presentation), we focused on the late timesteps (10-25) where such phenomena appeared. We verified whether different input components were aligned or misaligned with each other, therefore issuing a cancellation in the overall activities. To this end we computed the cosine similarity (scalar product) between all pairs of the three input components before and after label presentation (presentation vs processing phase \cref{fig:2}d). For positive data, the forward component was largely anti-aligned to both the backward and lateral components, suggesting that the decrease in activity was due to the bottom-up (forward) information flow canceling the top-down (backward) and recurrent (lateral) information flows (\cref{fig:2}d). Conversely, for negative data, the top-down and bottom-up information flows showed a higher degree of alignment, resulting in increased activations (surprise signal) (\cref{fig:2}d).

This analysis shows that our model reproduces hierarchical properties of predictive computations by generating information flows that result in surprise and cancellation signals. These signals are associated with the processing of negative and positive data, respectively, and involve distinct network information flows based on the dynamic cancellation of multiple input components. Although the degree of alignment across components could vary from instantiation to instantiation, these cancellation phenomena were highly robust.

\begin{figure}[h]
\centering
\includegraphics[width=\linewidth]{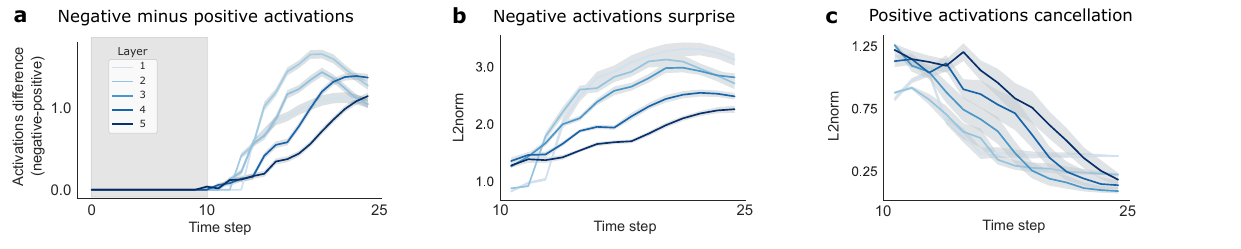}
\caption{{\footnotesize Activations surprise and cancellation order. All error bars are one sigma. (a) The negative minus positive activations (differences) over time (x-axis) are shown as a measure of the negative activation surprise signal, offset from the baseline of our positive activations. (b) The L2 norm of negative activations (y-axis) is shown across time (x-axis), visualizing the cancellation cascade during the processing phase. (c) Same as panel b for the cancellation of positive activations.}}
\label{fig:3}
\end{figure}

\subsection{Dynamical emergence of surprise and cancellation signals}
\label{sec:dyn}
We next interrogated the temporal characteristics of the cancellation and surprise information flows. We began by plotting activation differences across all layers (as performed in \cref{fig:2}c) in \cref{fig:3}a. This demonstrated that the encoding of positive versus negative data diverged more rapidly between early layers compared to later layers. To confirm this, we analyzed activations for negative and positive data during the processing phase, after introducing the label. 
For negative data activity grew faster for earlier layers despite label information being fed from the top of the hierarchy (\cref{fig:3}b). In the case of positive data, early layer activations led the cancellation cascade by returning to a lower activation state, prior to late layers, which establishes a bottom-up cancellation ordering. We also analyzed the cosine similarity between activations of consecutive layers, for both positive and negative data (\cref{fig:SI_2}), confirming this cascade ordering respectively for surprise and cancellation signals. Together, the findings shown in fig.~\ref{fig:3}a-c indicate that alignment and anti-alignment dynamics across layer activations, leading to surprise and cancellation signals, originate in early layers despite the introduction of the label at the top of the hierarchy.

\subsection{Interpreting latent representations which drive cancellations}
\label{sec:popdyn}

\begin{figure}[h]
\centering
\includegraphics[width=\linewidth]{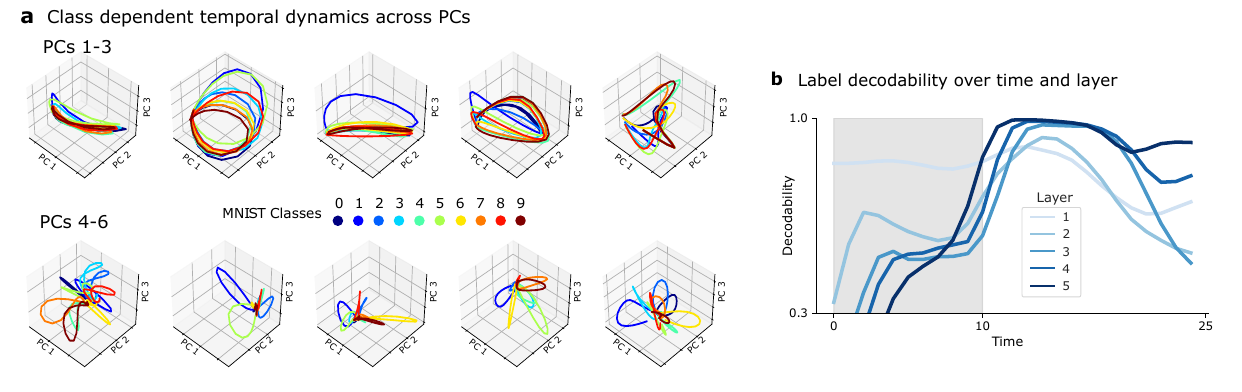}
\caption{{\footnotesize Latent representations and label decodability over both principal components and layer: (a) Representation of the layer-wise latent spaces on three dimensions via PCA where classes are represented by color. The first three PCs are shown to indicate a lack of class separability. Higher-order PCs are shown to indicate stronger class separability. (b) Decodability (y-axis) over time (x-axis) for different layers, indicating that the pre-label timesteps are driven by a distinct bottom-up temporal driving, and by contrast late processing time displays cancellation (and the associated drop in decodability) cascading from the bottom-up.}}
\label{fig:5}
\end{figure}

In order to understand the latent space mechanics driving cancellations on positive data, we sought to understand the intricate mechanics governing the latent space dynamics. We first plotted the average class-wise activations for various PCs in lower dimensions (\cref{fig:5}a). We observe that the lower-order PCs do not offer a strong representation of the class, but they do offer a consistent path through the space that starts and ends at the same point. This is in line with the mechanics of the network under positive data, which starts from an initially low activity and recovers to a similarly low activity following all the timesteps where the label is presented. In the higher-order PCs (4-6), chosen for their stronger representations, the same looping mechanics are shown. However, now the classes are represented in a separable manner. To further quantify this qualitative analysis we performed a decoding analysis highlighting the presence of label information across multiple PCs (\cref{fig:SI_3}c). For negative data, the looping behavior in the latent space does not occur, and the latent states drive away from the origin erratically in a class and label dependent manner. 

To measure the directionality of information flow throughout, 5 MLPs were trained on the latents of each of the 5 layer-wise activations for positive data (\cref{fig:5}b). High decodability indicates label-specific information. Two distinct cascades of decodability increase were observed: one for the presentation phase (timesteps 0-9, upon image presentation), and another for processing phase (timesteps 10-24, upon label introduction). These decodability rises during the presentation phase, and falls during the processing phase, reveal a layer-wise bottom-up temporal ordering. This bottom-up cascading response is consistent with the introduction of the image at the bottom of the layer hierarchy during the presentation phase, but opposite of the top-down label representation flow during the processing phase. Specifically, in the presentation phase, the decodability increases first for lower layers, indicating a bottom-up temporal ordering. By contrast, in the processing phase, the decodability decreases first for earlier layers despite a top-down injection of the label. Thus, even in the presence of top-down inputs, these results indicate that bottom-up cascades provide the dominant axis along which representational changes unfold over time.

\subsection{The Inverted Forward-Forward is a contrastive three-factor-learning rule which converges to synaptic drive cancellation}
\label{sec:heb}
In this section, we demonstrate an online learning variant, showing learned cancellation arising from minimization of a
contrastive softplus margin objective.

For a $N>3$ Forward-Forward architecture where $N$ is the total number of layers, $I$ is the data input, $\ell$ is the label input, \textit{W} is the recurrent lateral synaptic connection, \textit{F} is the bottom up synaptic connection, and \textit{B} is the top down synaptic connection, the dynamics of each layer are governed by
\begin{align*}
\dot {\vec{x_1}} &= \phi\left(W_1 \hat x_1 + F_1 I(t') + B_1 \hat x_2\right) \\
\dot {\vec{x_i}} &= \phi\left(W_i \hat x_i + F_i \hat x_{i-1} + B_i \hat x_{i+1}\right) \\
\dot {\vec{x_N}} &= \phi\left(W_N \hat x_N + F_N \hat x_{N-1} + B_N \ell(t')\right).
\end{align*}
The locally-defined loss for a one-step update takes the form of
\[
\mathcal{L}_i(t')
=
\sigma\!\left(
(-1)^{\eta(t')}
\left[
\mathbf{x}_i^\top(t')\mathbf{x}_i(t')-\theta
\right]
\right).
\]
where $\sigma(x)$ represents the softplus as $\sigma(x) = \log(1+ e^x)$ and is a smooth version of the ReLU nonlinearity. Importantly, the dynamics of the $\eta(t')$ are governed by a bistable dynamics:
\[
\eta(t')
=
1-\delta_{\ell(t'),c(I(t'))}.
\]

Here, $c(I(t'))$ denotes the class of the input image, and
$\delta_{a,b}$ is the Kronecker delta. Thus, $\eta(t')=0$ for a
matching input--label pair and $\eta(t')=1$ for a mismatched pair.
The binary variable $\eta(t')$ therefore determines the direction of
the local contrastive update.

% . Here $\delta_{ij}$ is the Kronecker delta notation, and $L(t')$ a signal that defines positive and negative data. This phenomenon could align with neuromodulator-induced shifts in underlying dynamics. It also suggests a criterion for selecting $\eta(t')$ based on the instantaneous surprise of the stimulus against the speculative label. Although beyond the scope of this work, the closure of this loop between activations and cost function may generate valuable insights into unsupervised variants of these learning rules \cite{ororbia_predictive_2023}.

At each processing timestep, we execute a local gradient-descent
update for each synaptic matrix:
\[
\partial_t W_i
=
-\alpha\nabla_{W_i}\mathcal{L}_i(t'),
\]
\[
\partial_t B_i
=
-\alpha\nabla_{B_i}\mathcal{L}_i(t'),
\]
and
\[
\partial_t F_i
=
-\alpha\nabla_{F_i}\mathcal{L}_i(t').
\]

By iteratively minimizing this locally-defined objective function, we seek a hierarchical structure that will work in concert with the other layers to minimize activations for positive data. For a mismatched input--label pair, minimizing the objective instead drives the layer activity above the threshold, thereby preventing cancellation and increasing the surprise signal.

Indeed this single-step update for a given layer takes the form of a three-factor Hebbian learning, since
\begin{align}
\mathbf{u}_i(t'-1)
&=
W_i\hat{\mathbf{x}}_i(t'-1)
+F_i\hat{\mathbf{x}}_{i-1}(t'-1)
+B_i\hat{\mathbf{x}}_{i+1}(t'-1),
\\[0.5em]
\nabla_{W_i}\mathcal{L}_i(t')
&=
2\underbrace{
(-1)^{\eta(t')}
\sigma'\!\left(
(-1)^{\eta(t')}
\left[
\mathbf{x}_i^\top(t')\mathbf{x}_i(t')-\theta
\right]
\right)
}_{\text{global gating factor}}
\\
&\quad{}\times
\underbrace{
\left[
\mathbf{x}_i(t')
\odot
\phi'\!\left(\mathbf{u}_i(t'-1)\right)
\right]
}_{\text{post-synaptic factor}}
\underbrace{
\hat{\mathbf{x}}_i^\top(t'-1)
}_{\text{pre-synaptic factor}}
\end{align}
where $\vec z(t'-1) = W_i \hat x_i(t'-1) + F_i \hat x_{i-1}(t'-1) + B_i \hat x_{i+1}(t'-1)$ is the input current into the nonlinearity. The factor of $2$ arises from differentiating
$\mathbf{x}_i^\top\mathbf{x}_i$ and may be absorbed into the
effective learning rate. 

This form of learning is formally a gated Hebbian or three-factor rule \cite{bahroun_normative_nodate, kusmierz_learning_2017, bredenberg_impression_2021, pogodin_kernelized_2020, portes_distinguishing_2022,bellec_solution_2020, murray_local_2019}  linking the locally-defined objective function to the product of the pre-synaptic current and the post-synaptic activation. The update is the product of a global scalar gate, a local
post-synaptic factor, and a local pre-synaptic factor, and therefore
takes the form of a three-factor Hebbian rule. 

These gradients have important implications on the shape of the learned solutions. Learned solutions where the gradient goes toward zero can occur under a number of conditions. These conditions include the direct cancellation of the synaptic drive currents (input components) governing the time dynamics of the hidden layer: $W_i \hat x_i + F_i \hat x_{i-1} + B_i \hat x_{i+1} = 0$. 

In this section, we established that the Inverted Forward-Forward
update has the form of gated three-factor Hebbian plasticity,
consisting of a binary sign factor, a class-dependent soft margin
gate, and local pre- and post-synaptic factors. The objective drives
positive-sample activity below the threshold and negative-sample
activity above it. Consequently, corrective gradients are strongest
when positive activity lies above the threshold and when negative
activity lies below the threshold. Cancellation of the lateral,
top-down, and bottom-up synaptic drives provides one stationary
solution of this local update rule.

\section{Discussion}

In this work, we have presented a biologically plausible mechanism that sheds light on the spatiotemporal and predictive nature of cortical processing without necessitating explicit predictions. Drawing inspiration from the Forward-Forward model, an emerging form of local, contrastive learning, 
we inverted its original objective function 
to reduce surprise activations for positive data.
This inversion incentivizes activity cancellation between information flows when top-down labels align with bottom-up sensory input. As a consequence, layers across the hierarchy develop the ability to predict and cancel each other's activities, facilitating the minimization of layer surprise.

These findings highlight the potential of simple, locally-defined learning principles to account for predictive properties similar to those observed in neocortical computations. Using a contrastive learning technique, we demonstrate how surprise and cancellation dynamics naturally arise, providing insights into neural processing. Importantly, these spatiotemporal predictions occur without an explicit prediction mechanism, suggesting an alternative approach to understanding neural computations.

\subsection{Biological plausibility}
\label{bioplaus}
The Inverted Forward-Forward model uses activation contrast to navigate credit assignment in hierarchical architectures in a bio-plausible fashion by incorporating: the absence of weight transport \cite{lillicrap_random_2016, portes_distinguishing_2022}, online compatible learning rules consistent with three-factor Hebbian plasticity, a biologically analogous separation of timescales, and the incorporation of structural hierarchy. First, feedback is separated from backpropagation and instead incorporated as a top-down signal avoiding weight transport. By disconnecting the $F$ and $B$ matrices, the flexible learning rule finds aligned but non-weight transported solutions.

Although global supervisory terms are sometimes downplayed as biologically implausible, it is worth comment that $\eta$ functions as a simple, singular global signal. In biological networks, diffusive small molecules can exert influence over a broad area via volume transmission, modulating the activity of many neurons beyond those directly connected by synapses. This global signal is analogous to the function of neuromodulators like acetylcholine, which can broadly signal state changes or attentional context, thereby gating plasticity across distributed neural populations without requiring fine-grained, point-to-point wiring \cite{bellec_solution_2020, picciotto2012acetylcholine}. This non-local signaling introduces biological plausibility, especially in the context of the local update rules of the Inverted Forward-Forward model, which involve local Hebbian plasticity gated by thresholded activation and signed by data type. The associated third factor ties an external signal, suggestive of the aforementioned neuromodulatory input, to the minimization (maximization) of layer activity for positive (negative) labels. The slow timescales of this switching, relative to both dynamics and plasticity, suggest a normative hypothesis for the role of perhaps overlooked small molecules \cite{kusmierz_learning_2017}. While we analogize $\eta$ to neuromodulation, we acknowledge that a globally broadcast, binary signal is a strong simplification of the noisy and complex dynamics of real neuromodulatory systems. This remains a key abstraction in our model but also represents an emerging hypothesis about the computational primitive underlying neuromodulatory dynamics.  

There is also a negative-free view, in which contrastive repulsion via the globally binary signal $\eta$ can be replaced by activity decorrelating regularizers, as in Latent Predictive Learning (LPL) \cite{halvagal_combination_2023}. This could serve to hold $\eta$ fixed to its positive-phase while offloading repulsion of representations to other known phenomena, such as homeostatic synaptic scaling, inhibitory–excitatory balance, and activity-dependent synaptic competition \cite{vogels2011inhibitory, turrigiano2004homeostatic, hua2004neural}.

Thus, the Inverted Forward-Forward model introduces biological plausibility to the direct competition of top-down and bottom-up signal processing, with intriguing implications for interpreting the hierarchy of biological systems \cite{siegle_survey_2021, garrett_stimulus_2023}. 

\section{Conclusion}
Taken together, our results demonstrate that predictive computations and surprise responses can emerge from local contrastive dynamics without explicit error signaling or weight transport. By offering a tractable, biologically grounded alternative to classical predictive coding, the Inverted Forward-Forward model provides a foundation for future explorations into how cortical hierarchies may implement efficient learning and inference through simple, distributed mechanisms.
% %\section{Conclusion}

\section*{Acknowledgments} We thank the Allen Institute founders, Paul G. Allen and Jody Allen, for their vision, encouragement and support. We acknowledge funding support from the Shanahan Family Foundation, the University of California Santa Cruz, the eSciences institute, the Allen Institute, and the University of Washington's Computational Neuroscience Center. The authors thank Matt Golub, Michael Buice, Lu Mi, Uygar Sumbul, Ryan Raut, Adrienne Fairhall, and Eric Shea-Brown for insightful discussions and supporting this work. We also acknowledge support from the eScience institute's partnership with Microsoft Azure credits for their support of some of the computational exploration in this work.

\newpage
\appendix

\section{Equivalence between Forward-Forward architecture and Hebbian learning}
\label{app:A}
In this section, we derive the local synaptic update produced by
minimizing the layer-wise objective independently at each processing
timestep.

For a $N>3$ Forward-Forward architecture where $N$ is the total number of layers, $I$ is the data input, and $\ell$ is the label input, the dynamics of each layer are governed by
\begin{align*}
\dot {\vec{x_1}} &= \phi\left(W_1 \hat x_1 + F_1 I(t') + B_1 \hat x_2\right) \\
\dot {\vec{x_i}} &= \phi\left(W_i \hat x_i + F_i \hat x_{i-1} + B_i \hat x_{i+1}\right) \\
\dot {\vec{x_N}} &= \phi\left(W_N \hat x_N + F_N \hat x_{N-1} + B_N \ell(t')\right).
\end{align*}

where $\hat{x_i} = \frac{\vec x_i}{|\vec x_i|}$ is the layer-normed pre-synaptic drive. 

The locally-defined loss for a one-step update takes the form of:

\[
\mathcal{L}_i(t')
=
\sigma\!\left(
(-1)^{\eta(t')}
\left[
\mathbf{x}_i^\top(t')\mathbf{x}_i(t')-T
\right]
\right)
\]

where $\sigma(x)$ represents the softplus as $\sigma(x) = log(1+ e^x)$ and is a smooth version of the ReLU nonlinearity. The binary variable $\eta(t')$ indicates whether the input--label
pair at timestep $t'$ is positive or negative:

\[
\eta(t')
=
1-\delta_{\ell(t'),c(I(t'))}.
\]

where $\delta_{ij}$ is the Kronecker delta notation. 

Here, $c(I(t'))$ denotes the class of the input image and
$\delta_{a,b}$ is the Kronecker delta. Therefore, $\eta(t')=0$ for
positive samples and $\eta(t')=1$ for negative samples. In the
reported implementation, this binary variable is assigned according
to the input--label pairing at each processing timestep.

This bistable switching-like dynamics could potentially occur with long timescales between switches and could have compelling correspondence with our understanding of neuromodulatory induced switching of the underlying dynamics. This also suggests a criterion for the selection of $\eta(t')$ on the instantaneous surprise of the stimulus against the speculative label. While beyond the scope of this work, the closure of this loop between activations and cost function may generate valuable insights into unsupervised variants of these learning rules.

We then execute a single step-gradient update in each parameter:
$$
\partial_t W_i = -\alpha \nabla_{W_{i}} \mathcal{L}(t')
$$
$$
\partial_t B_i = -\alpha \nabla_{B_{i}} \mathcal{L}(t')
$$
$$
\partial_t F_i = -\alpha \nabla_{F_{i}} \mathcal{L}(t')
$$

By iteratively minimizing this locally-defined objective function, we seek a hierarchical structure which will work in concert with the other layers to minimize activations for positive data. For a mismatched input--label pair, minimizing the objective instead drives the layer activity above the threshold, thereby preventing cancellation and increasing the surprise signal.

Indeed this single step update for a given layer takes the form of a Hebbian learning rule:

\begin{align}
\mathbf{u}_i(t'-1)
&=
W_i\hat{\mathbf{x}}_i(t'-1)
+F_i\hat{\mathbf{x}}_{i-1}(t'-1)
+B_i\hat{\mathbf{x}}_{i+1}(t'-1),
\\[0.5em]
\nabla_{W_i}\mathcal{L}_i(t')
&=
2\underbrace{
(-1)^{\eta}
\sigma'\!\left(
(-1)^{\eta}
\left[
\mathbf{x}_i^\top(t')\mathbf{x}_i(t')-T
\right]
\right)
}_{\text{global gating factor}}
\\
&\quad{}\times
\underbrace{
\left[
\mathbf{x}_i(t')
\odot
\phi'\!\left(\mathbf{u}_i(t'-1)\right)
\right]
}_{\text{post-synaptic factor}}
\underbrace{
\hat{\mathbf{x}}_i^\top(t'-1)
}_{\text{pre-synaptic factor}} .
\end{align}

where the factor of $2$ arises from differentiating
$\mathbf{x}_i^\top\mathbf{x}_i$ and may be absorbed into the learning
rate $\alpha$.

This takes the form of a gated Hebbian or three-factor rule \cite{bahroun_normative_nodate, kusmierz_learning_2017, bredenberg_impression_2021, pogodin_kernelized_2020}  linking the locally-defined objective function to the product of the pre-synaptic current and the post-synaptic activation. These local gradient-descent updates are applied separately at each processing timestep $t'$.

These gradients have important implications on the shape of the learned solutions. Learned solutions where the gradient goes toward zero can occur under a number of conditions. These conditions include the direct cancellation of the synaptic drive currents (input components) governing the time dynamics of the hidden layer: $W_i x_i + F_i x_{i-1} + B_i x_{i+1} = 0$.

\subsection{Linearizing everything and reducing the dimension of the layers to one unit each}

Our goal in this subsection is to strictly apply two simplifying assumptions to show how the inverted FF objective function enforces cancellation in an easily understandable environment. The first assumption is that the dynamics are linear. The second assumption is that each layer is only represented by a single unit to simplify our view of cancellation. These assumptions enforce direct learned cancellation of top-down and bottom-up signals in these networks.

In the limit of linear dynamics of the underlying network and linear dynamics of the locally-defined objective function:
$$
\mathcal{L_i}(t') = \sum x_i^2 - T
$$
and
$$
x_i = W_i \hat x_i + B_i \hat x_{i+1} + F_i \hat x_{i-1}
$$

The gradients give rise to the simple learning dynamics of the form:
$$
\dot W_i = \alpha (-1)^{\eta} x_i(t+1) x_i(t)
$$
$$
\dot F_i = \alpha (-1)^{\eta} x_i(t+1) x_{i-1}(t)
$$
$$
\dot B_i = \alpha (-1)^{\eta} x_i(t+1) x_{i+1}(t)
$$
With the edge cases taking the form of:
$$
\dot B_N \alpha (-1)^{\eta} x_i(t+1) L(t)
$$
$$
\dot F_1 \alpha (-1)^{\eta} x_1(t+1) I(t)
$$

These linear dynamics are determined by the discrete variable $\eta$ which tells you if the data is positive data or not. We therefore define positive data as where the labels are presented as classes and the images are presented as floats near the given class identity.

These nonlinear dynamics give us a rich learning sequence in which we have fast time dynamics governing the layer population scalars $x_i$ and the slower dynamics of the learned parameters. This separation of timescales allows us to represent the mean population activity over the trial to study the convergence of the learning dynamics to steady state solutions. 

\begin{figure}[h]
\centering

\includegraphics[width=\linewidth]{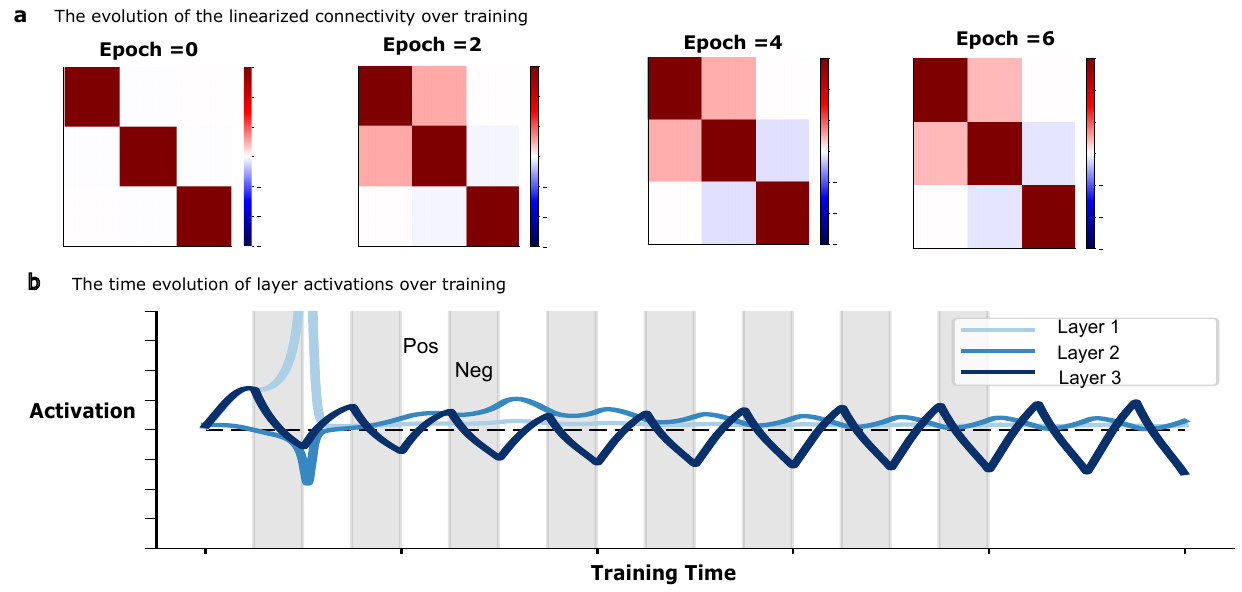}
\caption{Analysis of training dynamics and connectivity matrix in a Forward Forward network: (a) Over training, the evolution of the simplified connectivity matrix develops opposing terms resulting in the cancellation of matched signals (top-down and bottom-up) into layer 2. (b) The time course of training these linearized dynamics generates a system capable of switching between positive and negative data samples in an online fashion. Negative data is characterized by growing activations in layer 2 while positive data is characterized by cancelling activations in layer 2. These linearized dynamics provide a simplified playground to understand the emergence of cancellation with simple local learning rules.}
\label{fig:SI_1}
\end{figure}

The simulation of these linearized but still nonlinear dynamics for a three-layer network in an online learning setting confirms our analysis for this one-dimensional projection of population activity in that $F \rightarrow - B$ over training timesteps (\cref{fig:SI_1}). 

For a one-layer architecture with similar top-down/bottom-up representations of image class and label respectively, the positive data equivalence trivially forces the feedforward $F$ and feedback $B$ matrices to converge toward the negative of each other, $B \rightarrow - F$. This can be seen in the evolution of the above dynamics of the weight vectors. The only way that the linearized dynamics can go to zero is when $\dot B_1 = 0 = - x_i L(t)$ and thus since $L(t)$ is fixed at a non-zero value by the supervision, the $x_i$ must go to zero. To achieve this, we must have clean cancellation of the underlying dynamics in the hidden layer.  This forms the basis of the positive data cancellation and finds solutions which are consistent with increasing activation in response to mismatch or surprise.

\section{Temporal dynamics of layer similarity}

\begin{figure}[h]
\centering
\includegraphics[width=\linewidth]{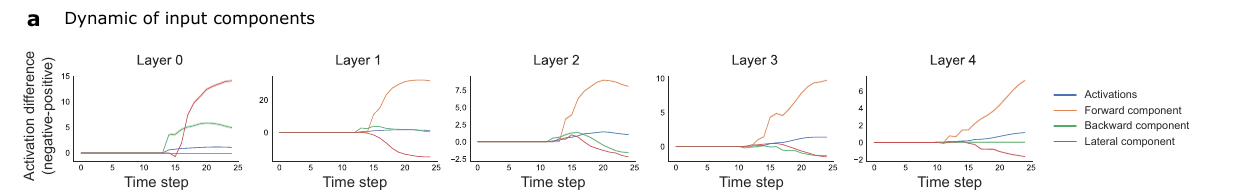}
\caption{Layerwise input components differences (negative minus positive) dynamics across the timesteps. Noticeably, the forward component is higher for negative data compared to positive data. For layer 0, the forward component is trivially identical for positive and negative data as it is driven by the input image.}
\label{fig:SI_2.0}
\end{figure}

\begin{figure}[h]
\centering
\includegraphics[width=\linewidth]{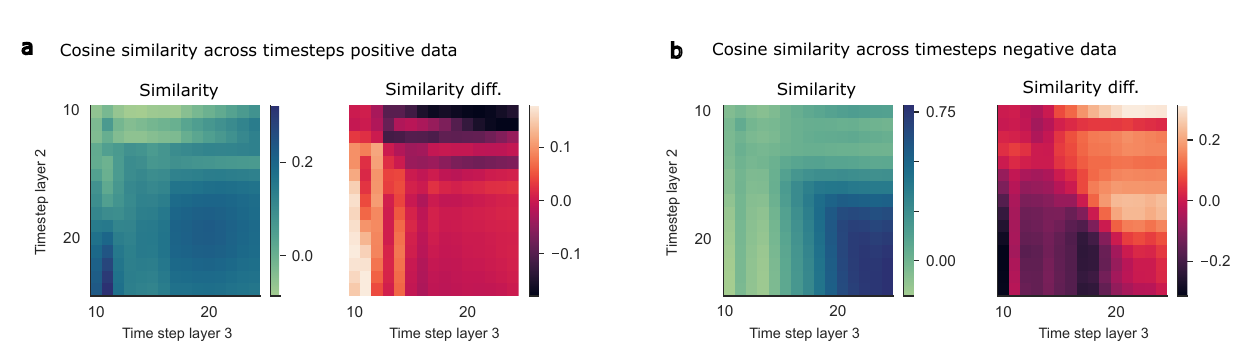}
\caption{Cosine similarity analysis for layer 2 and 3: (a) Across all timesteps of layer 2 and 3 (left panel). Similarity Difference metric (SD) across timesteps between layer 2 and 3. (b) Same analysis as panel (a) but for negative data.}
\label{fig:SI_2}
\end{figure}

We expanded the cosine similarity analysis across all timesteps for any two consecutive layers to further investigate this phenomenon. In \cref{fig:SI_2}a to \cref{fig:SI_2}b, we illustrate the cosine similarity between activations of layers 2 and 3 across any two timesteps during the processing phase for both positive (\cref{fig:SI_2}a left panel) and negative (\cref{fig:SI_2}b left panel) data. For positive data, cosine similarities decrease over timesteps confirming that activations across different layers decorrelate over this period. On the other hand, for negative data, similarities increase over the same period, confirming and generalizing our findings in the main text. This analysis demonstrates the emergence of a striking temporal ordering of cancellations (positive data) and activations (negative data) which reflects the structurally imposed hierarchy of the layers. This relationship emphasizes the importance of a mechanistic understanding going beyond the naive cancellation of image and label representations as their first collision. 

To examine the temporal dynamics further, we analyzed the difference between such similarities: for any pair of time steps, we computed the following metric. Denoted with $cos(a_{l2}(t_1),a_{l3}(t_2))$ is the cosine similarity between the activations $a_{l2}(t_1)$ of layer 2 at time $t_1$ and the activations $a_{l3}(t_2)$ of layer 3 at time $t_2$. We also defined the Similarity Difference $\textrm{SD}_{l23}(t_1, t_2)= cos(a_{l2}(t_1),a_{l3}(t_2))-cos(a_{l2}(t_2),a_{l3}(t_1))$. This quantity provides insight into the temporal dynamics because, for $t_1<t_2$, it is positive if the similarity between earlier activations in the first layer and later activations in the second layer is greater than the similarity between later activations in the first layer and earlier activations in the second layer. This value quantifies when current signals in one layer are analogous to subsequent signals in a second layer for any given timestep. A positive SD above the diagonal (accompanied by a negative SD below the diagonal) quantifies the influence of the first layer on subsequent timesteps in the second layer. This case, as described, is what we observed for negative data, confirming a bottom-up flow in late timesteps, \cref{fig:SI_2}b. For positive data, a top-down signal appears to flow into the layer for a few time steps before activities across layers decorrelate and cancellation of activity occurs, \cref{fig:SI_2}a right panel. 

This analysis validates the presence of two information flows for positive and negative data, with distinct temporal relationships between layers. It further illustrates that such information flows have specific dynamical properties across layers, where the activity in a given layer precedes or follows the activity in others across the hierarchy, enabling the generation of predictive types of signals.

\subsection{Visualizing the image specific low-dimensional dynamics}

\begin{figure}[h]
\centering
\includegraphics[width=\linewidth]{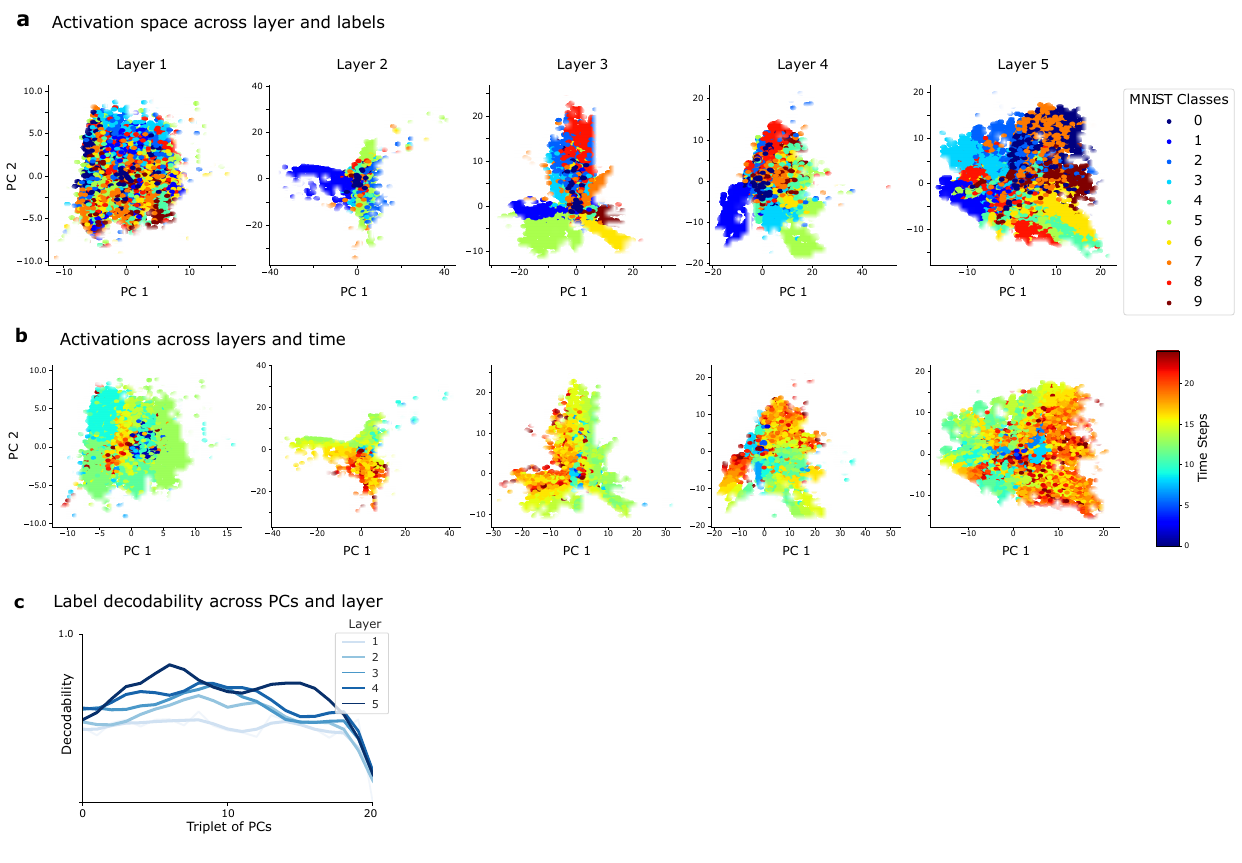}
\caption{Analysis of layer-wise latent spaces and decodability: (a) Representation of the layer-wise latent spaces on two dimensions via PCA where classes are represented by color. (b) Same latent space representation, but color-coded based on timestep. (c) Decodability (y-axis) across different layers' PCs (x-axis), indicating that PCs 4-10 capture rich representations. The x-axis label indicates the lowest PC out of the triplet used for decoding.}
\label{fig:SI_3}
\end{figure}

A temporal analysis of the same latent space in two dimensions was conducted on the higher-order PCs with stronger class representation (\cref{fig:SI_3}a). Activations start in the middle of the represented structure, before diverging and returning back to the beginning. This analysis shows behavior corroborating the above looping mechanics, driven by the recovery of the initial low activation state after initial excitement (\cref{fig:SI_3}b).  

\subsection{Label information and principal components}
To understand if the principal component representation of the network dynamics effectively captured the variance of the underlying data, we trained a slew of multilayer perceptrons (MLPs) on the latents of 1000 samples at every timestep, reduced in dimensionality by a sliding-window of three PCs. This analysis shows high label decodability for the PCs plotted above which showed cleaner separability (4-6) (\cref{fig:5}b), motivating the choice of these particular PCs. Additionally, most of the variance within this data is captured within the first 20 PCs, with decodability dropping to chance levels for PCs greater than 20.

\section{Predictive coding}

The hierarchical predictive dynamics analyzed this far, giving rise to surprise and cancellation signals, are specific of our model. We compare our model with established predictive coding networks (PCNs), first introduced by Rao \& Ballard \cite{rao_predictive_1999}, to further characterize these dynamics in contrast to those of PCNs. %, is pivotal in highlighting the novel contributions of the Forward-Forward network for modeling cortical computation. 
Predictive coding networks are characterized by a hierarchical structure wherein each layer predicts the subsequent layer's activity, informed by the product of its activity and a weight matrix, processed through a nonlinear function. The objective function of this predictive coding network is to minimize the loss:
\begin{equation}
\label{eq:2}
\mathcal{L}_{\textrm{layer}}(t) = | \phi(B\vec{x}_{l+1}) - \vec{x}_l|^2\,,
\end{equation}
which is often referred to as prediction error.
Here, \(x_l\) denotes the activity of layer \(l\), \(B\) represents the weight matrix, and \(\phi\) is a nonlinear activation function. The training process involves an alternating optimization strategy where the network first adjusts its weights to minimize the prediction error and subsequently refines the layer activations to further reduce the discrepancy between prediction and actual sensory input. This iterative process aims to model the brain's learning mechanism, which continually adapts to new information.

\begin{figure}[h]
\centering
\includegraphics[width=\linewidth]{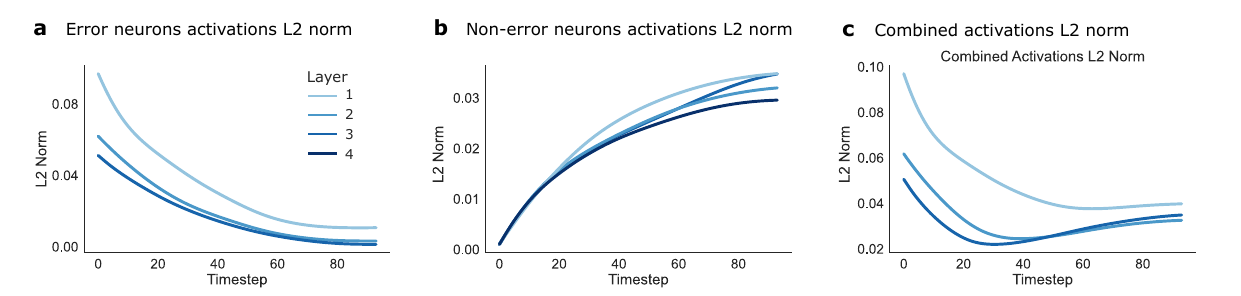}
\caption{Analysis of Predictive Coding Network: (a) The norm of activation of error neurons decreases with time for all layers. (b) Norm of activation of non-error neurons increases across timesteps with no relationship between layer amplitude and layer position. (c) Norm of compound activations of error and non-error neurons.}
\label{fig:10}
\end{figure}

The Forward-Forward architecture differs from this framework in its intrinsic generation of predictions. Rather than relying on a hand-coded error computation between layers with dedicated error neurons and prediction errors, the Forward-Forward network learns predictions through a local learning rule intrinsic to each layer. This critical difference generates a dynamic which is qualitatively different. \cref{fig:10}a to \cref{fig:10}c show respectively the activations of error neurons, non-error neurons, and the compound activity. None of the highlighted phenomena in the Forward-Forward dynamics is present in such a predictive coding model. Critically, in a PCN, there is no distinction between positive or negative data, no surprise or cancellation signals generated by the network, and there is no bottom-up (or top-down) cascade in the way information propagates through the networks. On the other hand, these elements are observed in cortical networks, and are naturally generated by the Forward-Forward model. 

The Forward-Forward's approach of eschewing hand-coded prediction errors leads to more biologically aligned phenomena, as it appears to reproduce the spatio-temporal bottom-up activity cascade observed in mice full field flash experiments, as highlighted by \cite{siegle_survey_2021}. This cascade did not appear in our implementation of a PCN, which instead demonstrated a rise in prediction errors across all layers simultaneously. Further research is needed towards understanding the conditions under which predictive coding networks (PCNs) might align with the Forward-Forward architecture's distinctive dynamics, challenging the boundaries of these computational models in cortical computation emulation.

In our predictive coding network (PCN), the inference and learning phases optimize the same following loss function:
$$
\mathcal{L}(t) = \sum_i \Vert\vec x_{i-1} - \phi(B_i \vec x_{i})\Vert_2^2, 
$$
where $x_0$ is clamped to the input image.
The inference phase then takes the form of minimizing the loss with respect to the neural activities for each layer $i$:
$$
\dot{\vec{x}}_{i} = \nabla_{\vec x_{i}} \mathcal{L}(t) = 2B_i^{\top}\frac{\dd \phi^\top}{\dd \vec{z}_{i}}\left(\vec x_{i-1} - \phi(B_i\vec x_i)\right) + \phi(B_{i+1}\vec x_{i+1}) - \vec x_{i},
$$
where $\vec z_i = B_i\vec x_i$.  This is equivalent to a leaky neuron subject to two sources of synaptic drive: 1) the feedback from top-down, and 2) the prediction error change.
Allowing this to evolve to convergence gives us $\vec x_i^{\star}$. 

The learning phase then adopts a gradient descent on the same loss with respect to $B_i$ for each layer $i$:
$$
\dot B_i = \nabla_{B_i} \mathcal{L} = \frac{\dd \phi^\top}{\dd \vec{z}_{i}^\star}(\vec x_{i-1}^\star - \phi(B_i\vec x_i^\star))(\vec x_i^\star)^\top,
$$
where we define $\vec z_i^\star = B\vec x_{i+1}^\star$ 
%+ F \vec x_{i-1}^\star + W \vec x_i^\star$
which is the convergence input current. Usefully, this one-step gradient also takes the form of a three-factor rule which combines pre-synaptic current, post-synaptic activity and a third gating or 'gain' factor. In this case, the third factor takes the form of the prediction error.

If we relax the convergence assumption of the 'inference phase' and simply conduct online learning on this cost function, we can directly compare these updates to the inverted-FF model through comparison of our three-factor terms. 

\subsection{Comparison between inverted FF, PCN and supervised update rules}

In the previous section we demonstrated that the update equations for the feedback weights $B$ evolve with a distinct third-factor responsible for gating the Hebbian updates of the weights in PCN. This form of the rule is notably quite different from the inverted FF. To round out our comparison to include a simple variant of supervised loss we include the third factor for a supervised loss. We choose a simplified variant (with no feedback nonlinearities) of random feedback to focus our attention on the form of supervisory error~\cite{lillicrap_random_2016}. 

These third factors for distinct learning rules can now be compared on the same standing:
$$
 \text{Third Factor} =
    \begin{cases}
       \sigma' ((-1)^{\eta}\vec x_i^T (t') \vec x_i (t') - T)(\vec x_i(t')) & \text{for the inverted FF}\\
%      \left(\phi(B\vec x_{i+1} + F \vec x_{i-1} + W \vec x_i) - \vec x_i\right) & \text{for the PCN}
    \left(\phi(B_i\vec x_{i+1}) - \vec x_i\right) & \text{for the PCN}\\
     \left(\prod_{j = N} ^iB_j(\phi(\vec x_{N}) - y^\star)\right) & \text{for a supervised signal}
    \end{cases} 
$$

We emphasize two primary differences between the inverted FF and the PCN third factors. 

The first is the presence of the contrastive sign flip designed to avoid the collapse of the dynamics onto the trivial solution. This contrastive term plays the role of the supervisory signal in which the information about the clamped label is passed to each layer through the top-down feedback and the global error signal (reminiscent of neuromodulator volume transmission) driving either the elimination or increase of the surprise signal.

The second chief difference is that the inverted FF conditions weight updates on surprise being above a threshold while the PCN conditions weight update upon the activity prediction error. This follows from the supervised case where in both models, the surprise (inverted FF) and the prediction error (PCN) are acting like error signals in the network gating the Hebbian pre-synaptic, post-synaptic coincidence update rules.

\subsection{Why is the PCN distinct from the inverted FF?}
\label{discussion1}

On the surface, the PCN and inverted FF are motivated by the same ambitions. They both avoid backpropogation in favor of local cost functions that admit three-factor descriptions of their learning. By using this unifying approach to focus on the third factor alone, we can appreciate the differences more clearly.
\begin{equation}
 \text{Third Factor} =
    \begin{cases}
       \sigma' ((-1)^{\eta}\vec x_i^T (t') \vec x_i (t') - T)(\vec x_i(t')) & \text{for the inverted FF}\\
%      \left(\phi(B\vec x_{i+1} + F \vec x_{i-1} + W \vec x_i) - \vec x_i\right) & \text{for the PCN}
    \left(\phi(B_i\vec x_{i+1}) - \vec x_i\right) & \text{for the PCN}
    \end{cases} 
\end{equation}
The inverted FF differs from the PCN in two key aspects: contrastive supervision and gating conditions. In the inverted FF, supervision involves clamping the label at the top and employing a contrastive signal. In contrast, even in the PCN with a clamped label, the contrastive signal is absent. The second distinction lies in the conditioning of weight updates—on layer activity in the inverted FF and prediction error in the PCN. These differences account for the variations in steady-state activity and ordering.

\section{Mechanisms behind cancellation order}
As a result of our simulation, a compelling non-trivial logic has emerged from the model's hierarchical predictive dynamics, which are often difficult to comprehend. Here, we seek to explain the fundamental mechanistic principles underlying the network's information flow generation. We determined that despite the fact that such insights are difficult to isolate or prove, they may still be necessary to comprehend the model's inner mechanisms.

For the initial presentation phase of both positive and negative data, the image representation flows from the bottom-up. The differences are, however, quite different in the processing phases. For negative data, the processing phase induces a top down signal carrying label information downward through all layers. This top-down label signal causes relatively small increases in layer activation magnitude. It is only when the label information reaches the bottom, that the activation response grows dramatically, indicating a mismatch and evoking a large and sustained excitatory surprise. For positive data, the label representation traverses to the bottom layer without inducing cancellations, whereby the cancellations then start in a bottom-up manner, despite the top-down label representation.

As the presentation phase blends with the processing phase, it is insightful to note that, neither for the positive nor negative case is some predisposed behavior taking place. The layers in the network do not amplify or reduce their activations until the label representation reaches at least one layer below them and sends a label-infused representation back upwards. This suggests the bottom-up cancellation could be a result of the network's optimizing drive to alter activities when in a familiar state relative to training. During training,  there is a continuous and consistent exposure to label-augmented activations, particularly past the early stages, which ingrains a behavior within the network. The network recognizes label saturated activations as the dominant trend it should ideally be prepared for. Given this recognition, the network is best equipped for cancellation when it encounters activation components (forward, backward, recurrent) with label information coming from all components, not just from the top. Under this concept, a layer lower in the hierarchy would be predisposed to cancel first, due to the lower number of potential layers below lacking label-infused information which would thereby block cancellation. Thus, cancellation in lower layers would be followed by a transmission of label-infused activities upwards to the next layer to induce subsequent phases of bottom-up cancellation.

\subsection{ Capturing these cancellation dynamics in a toy amplitude, orientation model}

We present a phenomenological model of these cancellation dynamics. Importantly, this simple model is highly portable to new situations and ordering of image presentation, and label clamp. To build this toy model, we start with the following assumptions:
\begin{itemize}
    \item We have 5 layers whose dynamics are connected via a hierarchical chain of Feedforward, $F$, Feedback $B$ and recurrent, $W$ coupling.
    \item Each layer is represented by two neuronal degrees of freedom. Importantly, this allows us to represent the activation of each layer as an Amplitude, $A_i$ and an orientation $\phi_i$ via a polar coordinate transformation. 
    \item The image input layer is represented by a fixed amplitude and an angle which is clamped to image identity. (e.g. 0 for image 0 and $90^o$ from image 1). Note, the capacity of these two layer networks is a single-bit of information, whether the input image is 0 or $90^o$.
    \item The label is coupled into the top layer and is either governed as a clamping process during training or during inference as low-pass filter of the inferred orientation of the input image. 
\end{itemize}

Combining these assumptions into an inverted Forward-Forward inspired loss function, we propose:
$$
\mathcal{L} \approx \sum_{i \in layers}\underbrace{\left[2 cos(\phi_{i-1} - \phi_{i+1}) -1\right]}_{\text{Is this positive data?}} \overbrace{\sigma ( A_i - T)}^{\text{Saturated Over Threshold}}
$$

Taking the gradient of this loss with respect to the degrees of freedom tells us the dynamics along the chain:
$$
\theta_i \equiv \phi_{i-1} - \phi_{i+1}
$$
$$
\dot{A_i} = \left(1-cos(\theta_i) - A_i\right)
$$
$$
\dot{\phi_i} = A_{i-1} sin(\phi_{i-1} - \phi_i) + A_{i+1} sin(\phi_{i+1} - \phi_i)
$$

This suggests that the dynamics of information being passed along this hierarchically arranged chain will take two forms, (i) a set of propagation equations along the amplitude and (ii) a set of propagation equations of the orientation.

Conceptually, the dynamics of amplitude and orientation are very coupled. To get above threshold amplitude you must have disagreement of the input angles F and B. To get change to the orientation you must have non-zero amplitude. The dynamics of this chain can be shown to propagate in the following way:
\begin{enumerate}
    \item Nonzero disagreement between angles results in growing surprise/amplitude
    \item Each layer's orientation is pulled into agreement with the layers above and below. The strength of that pull is mediated by the amplitude of each layer.
    \item With growing alignment, the amplitude of each layer decays back toward zero. 
\end{enumerate}

\begin{figure}[H]
\centering
\includegraphics[width=0.5\linewidth]{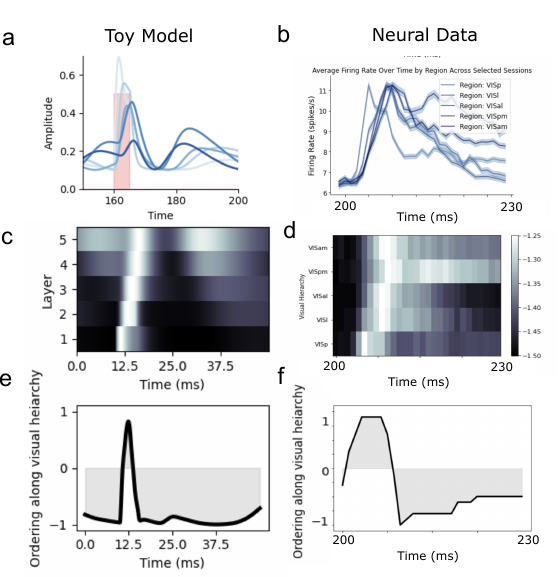}
\caption{Comparison of Neural Data and Toy cancellation model dynamics: (a) A temporal response of the toy model to a flash signified by the red illumination. Layers are colored in descending saturation of blue. (b) A comparable sequence of flash-on (presented in the Allen Institute's Visual Coding Neuropixels dataset) ordered from V1Sp to V1Sam consistent with the hierarchy of \citep{siegle_survey_2021}. (c-d) The same dynamics in the toy model and neural data results depicting the L2 norm of the layer on a log scale. (e-f) The ordered Spearman correlation of the amplitude against the hierarchy location. Negative values indicate that layers lower in the hierarchy tend to have lower amplitudes.. The initialization of the flash in both the toy model and the neural data show a dramatic inversion of this measure as the wave propagates upward through the layers before experiencing cancellation. We use the similarity between the amplitude-orientation dynamics f the toy model and the sequence observed in data and the FF model to suggest a connection between these three systems in mechanism of cancellation.}
\label{fig:SI_6}
\end{figure}

We explore these dynamics numerically (\cref{fig:SI_6}), in a situation similar to the flash cancellation observed in the Visual Coding Neuropixels dataset. 

\section{Visual Coding Neuropixels dataset analysis}

\subsection{Hierarchical neural activity cascades reproduce in neural data}
\label{sec:neuraldata}

\begin{figure}[h]
\centering
\includegraphics[width=\linewidth]{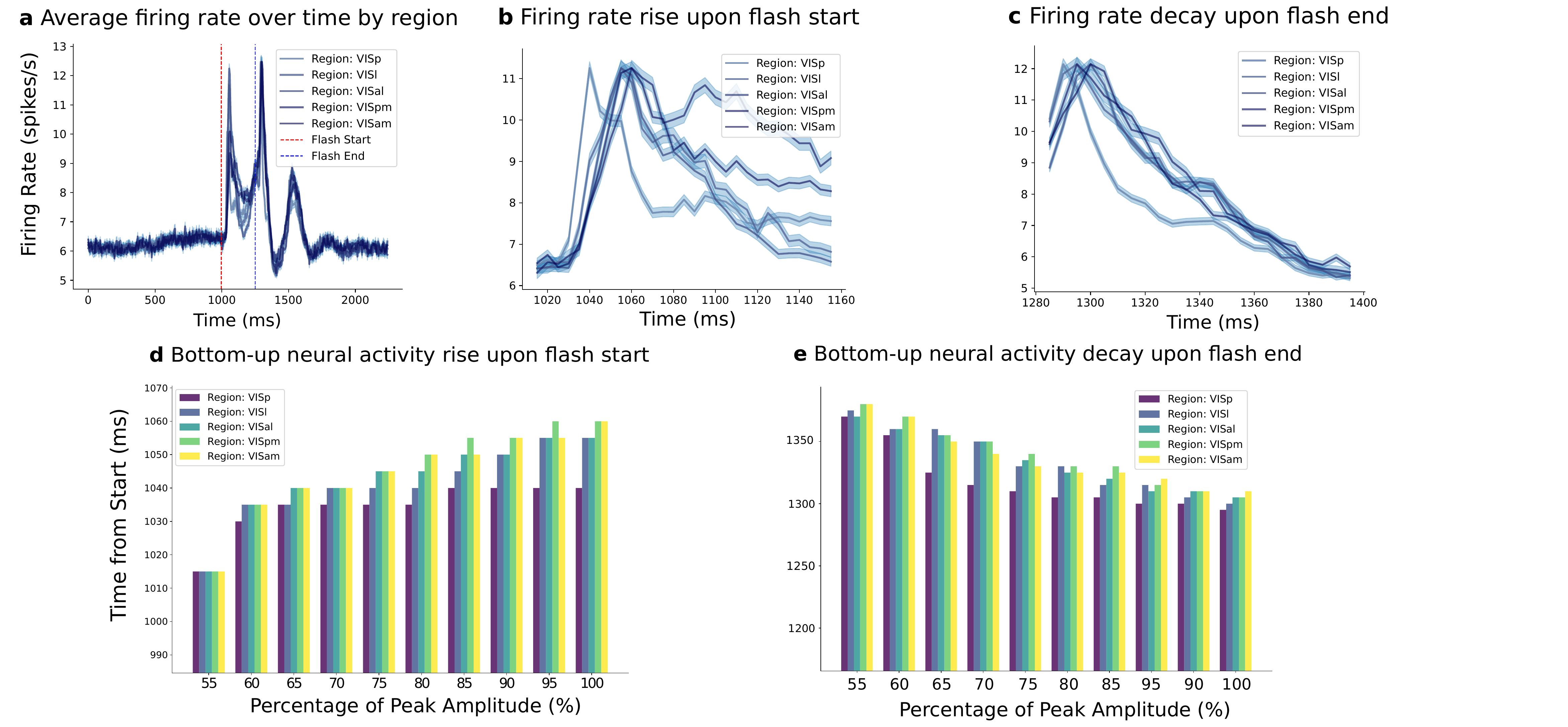}
\caption{{\footnotesize Average neural firing rate recordings resultant from full field flashes in mice. This data was sourced from the Visual Coding Neuropixels dataset provided by the Allen Institute. All error bars are one sigma. (a) Neural firing rate recordings averaged over many flashes across many flash-exposure sessions. (b) The neural activity rise upon flash onset presents a bottom-up layer-wise rising cascade. (c) The neural activity fall upon flash offset presents a bottom-up falling cascade. (d) The layer-wise neural activity rise upon flash onset takes on a bottom-up hierarchical cascade (read left to right). (e) The layer-wise neural activity fall upon flash offset immediately presents a bottom-up decay pattern starting from peak amplitude (read right to left). }}
\label{fig:4}
\end{figure}

Prior work has introduced the concept of hierarchies in the visual cortex and a cascading bottom-up information flow \cite{siegle_survey_2021}. In order to draw a direct comparison, we analyzed data from the Allen Institute's Visual Coding Neuropixels dataset, specifically looking at full-field flashes in mice. These experimental sessions capture region-specific neural data from mice that are exposed to a set of visual stimuli. Our analysis in \cref{fig:4} aggregates data across 50 experimental sessions where mice are exposed to a full-field flash, capturing hierarchical region data from 5 areas of the visual cortex. We average all neural activity around the stimuli and normalize it for different region-specific baselines to obtain insight into the average neural response to a single flash (see \cref{app:visual_details}).

The task paradigms are not identical: the model is trained on semantic mismatch, whereas the neural data are recorded during sensory transients evoked by full-field flashes. Nevertheless, both setups involve stimuli that propagate through hierarchical circuits. This motivates a comparison to assess whether the model’s emergent cascade of suppression resembles the hierarchical dynamics measured in vivo. The flash-evoked dataset represents the most suitable publicly available approximation to our setup, despite not being designed to probe predictive computations. Because the paradigm recruits the same hierarchical structures, the transient responses can serve as a proxy for mismatch-related dynamics. This comparison enables evaluation of the model’s predictions and motivates future experiments to directly test whether hierarchical suppression arises as the model suggests.

Our analysis focuses on the activity rise when the stimulus is presented and the activity fall when the stimulus is turned off  (\cref{fig:4}a and \cref{fig:4}b), as the neural activity pattern between stimulus onset and offset is dominated by transients. Our analysis in \cref{fig:4}d indicates that neural activity takes on a bottom-up layer-wise temporal cascade at the very beginning of the neural response to flash onset. This bottom-up temporal cascade is also consistent in the neural response to the flash offset, whereby the neural activity decays first in early layers in the visual hierarchy relative to later layers (\cref{fig:4}e).

These bottom-up layer-wise responses to changing stimuli are reminiscent of the Inverted Forward-Forward model's own bottom-up layer-wise response patterns. While the experimental design of the Inverted Forward-Forward model involves a single stimulus (label presentation during the processing phase), it does not directly parallel the stimulus offset observed in the flash experiments. However, the similarity in the bottom-up cascade of neural activity suggests that the Inverted Forward-Forward model captures some of the hierarchical processing dynamics observed in the visual cortex.

In both the visual cortex and the Inverted Forward-Forward model, high-level regions or representations integrate sensory inputs over time to form expectations. In the visual cortex, these high-level regions may generate prediction errors in response to new stimuli, even when the stimuli are expected, as seen with the flash. Similarly, the Inverted Forward-Forward model shows a bump in activity during the label presentation, even for positive data where it is trained to cancel such activity. This bump likely reflects the model's response to a salient stimulus before the learned cancellation mechanisms reduce the activity. This parallel suggests that the mechanisms underlying the Inverted Forward-Forward model may offer insights into the spatiotemporal cancellation observed in neural data. The model's ability to reproduce hierarchical and temporal processing patterns sheds light on how high-level expectations and prediction errors shape neural activity across different layers, mirroring some of the observed dynamics in the visual cortex.

Experimentally, the Inverted Forward-Forward model brings at least two testable predictions:

\begin{itemize}
    % \item Prediction 1 (Top-down disruption): The Inverted Forward-Forward predicts that the selective inactivation of top-down pathways in a sensory hierarchy should more severely impair the learned suppression of expected stimuli than the initial excitatory response to novel or unexpected stimuli. 

   \item Prediction 1 (Learned-suppression disruption): The Inverted Forward-Forward predicts that if the third factor $\eta$ is disrupted, this will impair the learned suppression response for a novel stimulus. Perhaps disrupting a novelty-signaling neuromodulator (e.g., via optogenetic inactivation of cholinergic inputs) during a predictive task would impair the learned suppression of expected stimuli without affecting the initial response to unexpected ones.

\item Prediction 2 (Ordered suppression across the hierarchy): 
The IFF model not only accounts for the cascade of stimulus-evoked responses across visual areas, but also predicts that suppression itself should follow the same bottom-up ordering. Once a stimulus–outcome association is well learned, predictable stimuli should evoke progressively reduced activity, appearing earlier in lower-order visual areas (e.g., V1) than in higher-order ones. Whether the existing flash responses are already the expression of this learned suppression or whether they are just ordinary evoked responses, the model makes a concrete prediction: suppression dynamics should cascade bottom-up through the hierarchy rather than emerging in a different fashion.
\end{itemize}

\subsection{Further Details: Visual Neuropixels Dataset Analysis}
\label{app:visual_details}

The Visual Neuropixels dataset comprises recordings of neural activity from the visual cortex of mice in response to a standardized battery of visual stimuli. This dataset was generated using Neuropixels probes, which offer high temporal resolution and the ability to record from multiple brain regions simultaneously.

We aggregated data from 50 sessions, overlaying neural data from each unit across all sessions for a fixed window of 2500 ms around each flash stimulus. This allowed us to examine the neural responses consistently over a significant temporal window.

To ensure comparability across regions, we normalized the region-wise average firing rate by the number of units recorded in each region. This step was crucial to account for differences in the number of recorded units across various brain areas.

Further normalization was performed to align all baselines to the average baseline firing rate and all peaks to the average peak firing rate. The normalization process was mathematically represented as follows:

\[
\tilde{R}_{\text{region}} = \left( R_{\text{region}} - B_{\text{region}} \right) \cdot \frac{P_{\text{avg}} - B_{\text{avg}}}{P_{\text{region}} - B_{\text{region}}} + B_{\text{avg}}
\]

where:
\begin{itemize}
    \item \( \tilde{R}_{\text{region}} \) is the normalized neural response for a specific region.
    \item \( R_{\text{region}} \) is the raw neural response for the region.
    \item \( B_{\text{region}} \) is the baseline firing rate for the region.
    \item \( P_{\text{region}} \) is the peak firing rate for the region.
    \item \( B_{\text{avg}} \) is the average baseline firing rate across all regions.
    \item \( P_{\text{avg}} \) is the average peak firing rate across all regions.
\end{itemize}

This procedure allowed us to standardize the neural responses, facilitating meaningful comparisons across different conditions and sessions. 

This comprehensive normalization process enabled a robust analysis of the neural coding mechanisms in response to visual stimuli, leveraging the extensive data provided by the Visual Neuropixels dataset.

% At the end of the manuscript:
% \bibliography{references}
\bibliographystyle{unsrtnat}
\bibliography{biblio}

@article{bassett_network_2017,
	title = {Network neuroscience},
	volume = {20},
	url = {https://www.nature.com/articles/nn.4502},
	number = {3},
	urldate = {2023-09-27},
	journal = {Nature neuroscience},
	author = {Bassett, Danielle S. and Sporns, Olaf},
	year = {2017},
	note = {Publisher: Nature Publishing Group US New York},
	pages = {353--364},
}

@article{siegle_survey_2021,
	title = {Survey of spiking in the mouse visual system reveals functional hierarchy},
	volume = {592},
	url = {https://www.nature.com/articles/s41586-020-03171-x},
	number = {7852},
	urldate = {2023-09-27},
	journal = {Nature},
	author = {Siegle, Joshua H. and Jia, Xiaoxuan and Durand, Séverine and Gale, Sam and Bennett, Corbett and Graddis, Nile and Heller, Greggory and Ramirez, Tamina K. and Choi, Hannah and Luviano, Jennifer A. and others},
	year = {2021},
	note = {Publisher: Nature Publishing Group UK London},
	pages = {86--92},
}

@article{chaudhuri_large-scale_2015,
	title = {A large-scale circuit mechanism for hierarchical dynamical processing in the primate cortex},
	volume = {88},
	url = {https://www.cell.com/neuron/pdf/S0896-6273(15)00765-5.pdf},
	number = {2},
	urldate = {2023-09-27},
	journal = {Neuron},
	author = {Chaudhuri, Rishidev and Knoblauch, Kenneth and Gariel, Marie-Alice and Kennedy, Henry and Wang, Xiao-Jing},
	year = {2015},
	note = {Publisher: Elsevier},
	pages = {419--431},
}

@misc{garrett_stimulus_2023,
	title = {Stimulus novelty uncovers coding diversity in visual cortical circuits},
	copyright = {© 2023, Posted by Cold Spring Harbor Laboratory. This pre-print is available under a Creative Commons License (Attribution-NonCommercial-NoDerivs 4.0 International), CC BY-NC-ND 4.0, as described at http://creativecommons.org/licenses/by-nc-nd/4.0/},
	url = {https://www.biorxiv.org/content/10.1101/2023.02.14.528085v2},
	doi = {10.1101/2023.02.14.528085},
	language = {en},
	urldate = {2023-09-27},
	publisher = {bioRxiv},
	author = {Garrett, Marina and Groblewski, Peter and Piet, Alex and Ollerenshaw, Doug and Najafi, Farzaneh and Yavorska, Iryna and Amster, Adam and Bennett, Corbett and Buice, Michael and Caldejon, Shiella and Casal, Linzy and D’Orazi, Florence and Daniel, Scott and Vries, Saskia EJ de and Kapner, Daniel and Kiggins, Justin and Lecoq, Jerome and Ledochowitsch, Peter and Manavi, Sahar and Mei, Nicholas and Morrison, Christopher B. and Naylor, Sarah and Orlova, Natalia and Perkins, Jed and Ponvert, Nick and Roll, Clark and Seid, Sam and Williams, Derric and Williford, Allison and Ahmed, Ruweida and Amine, Daniel and Billeh, Yazan and Bowman, Chris and Cain, Nicholas and Cho, Andrew and Dawe, Tim and Departee, Max and Desoto, Marie and Feng, David and Gale, Sam and Gelfand, Emily and Gradis, Nile and Grasso, Conor and Hancock, Nicole and Hu, Brian and Hytnen, Ross and Jia, Xiaoxuan and Johnson, Tye and Kato, India and Kivikas, Sara and Kuan, Leonard and L’Heureux, Quinn and Lambert, Sophie and Leon, Arielle and Liang, Elizabeth and Long, Fuhui and Mace, Kyla and Abril, Ildefons Magrans de and Mochizuki, Chris and Nayan, Chelsea and North, Katherine and Ng, Lydia and Ocker, Gabriel Koch and Oliver, Michael and Rhoads, Paul and Ronellenfitch, Kara and Schelonka, Kathryn and Sevigny, Josh and Sullivan, David and Sutton, Ben and Swapp, Jackie and Nguyen, Thuyanh K. and Waughman, Xana and Wilkes, Joshua and Wang, Michael and Farrell, Colin and Wakeman, Wayne and Zeng, Hongkui and Phillips, John and Mihalas, Stefan and Arkhipov, Anton and Koch, Christof and Olsen, Shawn R.},
	month = feb,
	year = {2023},
	note = {Pages: 2023.02.14.528085
Section: New Results},
}

@article{piet_behavioral_2023,
	title = {Behavioral strategy shapes activation of the {Vip}-{Sst} disinhibitory circuit in visual cortex},
	url = {https://www.biorxiv.org/content/10.1101/2023.04.28.538575.abstract},
	urldate = {2023-09-27},
	journal = {bioRxiv},
	author = {Piet, Alex and Ponvert, Nick and Ollerenshaw, Douglas and Garrett, Marina and Groblewski, Peter A. and Olsen, Shawn and Koch, Christof and Arkhipov, Anton},
	year = {2023},
	note = {Publisher: Cold Spring Harbor Laboratory},
	pages = {2023--04},
}

@article{badre_frontal_2018,
	title = {Frontal cortex and the hierarchical control of behavior},
	volume = {22},
	url = {https://www.cell.com/trends/cognitive-sciences/fulltext/S1364-6613%2817%2930245-0?elsca1=etoc&amp%3Belsca2=email&amp%3Belsca3=1364-6613_201802_22_2_&amp%3Belsca4=Cell+Press&code=cell-site},
	number = {2},
	urldate = {2023-09-27},
	journal = {Trends in cognitive sciences},
	author = {Badre, David and Nee, Derek Evan},
	year = {2018},
	note = {Publisher: Elsevier},
	pages = {170--188},
}

@article{froudarakis_visual_2019,
	title = {The {Visual} {Cortex} in {Context}},
	volume = {5},
	issn = {2374-4642, 2374-4650},
	url = {https://www.annualreviews.org/doi/10.1146/annurev-vision-091517-034407},
	doi = {10.1146/annurev-vision-091517-034407},
	language = {en},
	number = {1},
	urldate = {2023-09-27},
	journal = {Annual Review of Vision Science},
	author = {Froudarakis, Emmanouil and Fahey, Paul G. and Reimer, Jacob and Smirnakis, Stelios M. and Tehovnik, Edward J. and Tolias, Andreas S.},
	month = sep,
	year = {2019},
	pages = {317--339},
}

@article{khan_contextual_2018,
	title = {Contextual signals in visual cortex},
	volume = {52},
	url = {https://www.sciencedirect.com/science/article/pii/S0959438818300825?casa_token=xIXteX3UtOgAAAAA:jgedYEVdm7FxHu6zV8-YxvKg_7nKGt73wzlHmrNIunlgUgKFzvmP16Socgq6cQALCeddYX6vA_U_},
	urldate = {2023-09-27},
	journal = {Current Opinion in Neurobiology},
	author = {Khan, Adil G. and Hofer, Sonja B.},
	year = {2018},
	note = {Publisher: Elsevier},
	pages = {131--138},
}

@article{rao_predictive_1999,
	title = {Predictive coding in the visual cortex: a functional interpretation of some extra-classical receptive-field effects},
	volume = {2},
	shorttitle = {Predictive coding in the visual cortex},
	url = {https://www.nature.com/articles/nn0199_79},
	number = {1},
	urldate = {2023-09-27},
	journal = {Nature neuroscience},
	author = {Rao, Rajesh PN and Ballard, D.H.},
	year = {1999},
	note = {Publisher: Nature Publishing Group},
	pages = {79--87},
}

@article{rao_predictive_2022,
  title = {Dynamic predictive coding: A model of hierarchical sequence learning and prediction in the neocortex},
  volume = {20},
  ISSN = {1553-7358},
  url = {http://dx.doi.org/10.1371/journal.pcbi.1011801},
  DOI = {10.1371/journal.pcbi.1011801},
  number = {2},
  journal = {PLOS Computational Biology},
  publisher = {Public Library of Science (PLoS)},
  author = {Jiang,  Linxing Preston and Rao,  Rajesh P. N.},
  editor = {Rubin,  Jonathan},
  year = {2024},
  month = feb,
  pages = {e1011801}
}

@article{wolpert_internal_1998,
	title = {Internal models in the cerebellum},
	volume = {2},
	url = {https://www.cell.com/trends/cognitive-sciences/fulltext/S1364-6613(98)01221-2},
	number = {9},
	urldate = {2023-09-27},
	journal = {Trends in cognitive sciences},
	author = {Wolpert, Daniel M. and Miall, R.C. and Kawato, Mitsuo},
	year = {1998},
	note = {Publisher: Elsevier},
	pages = {338--347},
}

@book{schenck_adaptive_2008,
	title = {Adaptive internal models for motor control and visual prediction},
	url = {https://www.google.com/books?hl=en&lr=&id=aorWUm2hbLYC&oi=fnd&pg=PA1&dq=schenck+internal+model&ots=3gmKIJ1I3M&sig=2RzLtJQ1QM0_kMsORSD7AKpUNqs},
	number = {20},
	urldate = {2023-09-27},
	publisher = {Logos Verlag Berlin GmbH},
	author = {Schenck, Wolfram},
	year = {2008},
}

@article{kawato_internal_1999,
	title = {Internal models for motor control and trajectory planning},
	volume = {9},
	url = {https://www.sciencedirect.com/science/article/pii/S0959438899000288?casa_token=4-yYDedK9_8AAAAA:JagChpbOID-Lg21iEfnobU0W1yZHGx_eyZ79KtM5K6j-NkoL3z69DSVRrmgzUhPvLg1aNNmzHsNj},
	number = {6},
	urldate = {2023-09-27},
	journal = {Current opinion in neurobiology},
	author = {Kawato, Mitsuo},
	year = {1999},
	note = {Publisher: Elsevier},
	pages = {718--727},
}

@article{mechelli_where_2004,
	title = {Where bottom-up meets top-down: neuronal interactions during perception and imagery},
	volume = {14},
	shorttitle = {Where bottom-up meets top-down},
	url = {https://academic.oup.com/cercor/article-abstract/14/11/1256/331439},
	number = {11},
	urldate = {2023-09-27},
	journal = {Cerebral cortex},
	author = {Mechelli, Andrea and Price, Cathy J. and Friston, Karl J. and Ishai, Alumit},
	year = {2004},
	note = {Publisher: Oxford University Press},
	pages = {1256--1265},
}

@misc{hinton_forward-forward_2022,
	title = {The {Forward}-{Forward} {Algorithm}: {Some} {Preliminary} {Investigations}},
	shorttitle = {The {Forward}-{Forward} {Algorithm}},
	url = {http://arxiv.org/abs/2212.13345},
	urldate = {2023-09-27},
	publisher = {arXiv},
	author = {Hinton, Geoffrey},
	month = dec,
	year = {2022},
	note = {arXiv:2212.13345 [cs]},
}

@inproceedings{bredenberg_impression_2021,
	title = {Impression learning: {Online} representation learning with synaptic plasticity},
	shorttitle = {Impression learning},
	url = {https://openreview.net/forum?id=MAorPaLqam_},
	language = {en},
	urldate = {2023-09-28},
	author = {Bredenberg, Colin and Lyo, Benjamin S. H. and Simoncelli, Eero P. and Savin, Cristina},
	month = nov,
	year = {2021},
}

@inproceedings{pogodin_kernelized_2020,
	title = {Kernelized information bottleneck leads to biologically plausible 3-factor {Hebbian} learning in deep networks},
	volume = {33},
	url = {https://proceedings.neurips.cc/paper/2020/hash/517f24c02e620d5a4dac1db388664a63-Abstract.html},
	urldate = {2023-09-28},
	booktitle = {Advances in {Neural} {Information} {Processing} {Systems}},
	publisher = {Curran Associates, Inc.},
	author = {Pogodin, Roman and Latham, Peter},
	year = {2020},
	pages = {7296--7307},
}

@article{bahroun_normative_nodate,
	title = {A {Normative} and {Biologically} {Plausible} {Algorithm} for {Independent} {Component} {Analysis}},
	language = {en},
	author = {Bahroun, Yanis and Chklovskii, Dmitri B and Sengupta, Anirvan M},
}

@article{kusmierz_learning_2017,
	series = {Computational {Neuroscience}},
	title = {Learning with three factors: modulating {Hebbian} plasticity with errors},
	volume = {46},
	issn = {0959-4388},
	shorttitle = {Learning with three factors},
	url = {https://www.sciencedirect.com/science/article/pii/S0959438817300612},
	doi = {10.1016/j.conb.2017.08.020},
	urldate = {2023-09-28},
	journal = {Current Opinion in Neurobiology},
	author = {Kuśmierz, Łukasz and Isomura, Takuya and Toyoizumi, Taro},
	month = oct,
	year = {2017},
	pages = {170--177},
}

@article{Gilbert_Li_2013, title={Top-down influences on visual processing}, volume={14}, rights={2013 Springer Nature Limited}, ISSN={1471-0048}, DOI={10.1038/nrn3476}, abstractNote={In contrast to the traditional idea that the processing of visual information consists of a sequence of feedforward operations, with neuronal functional properties taking on increasing complexity as the information progresses through a hierarchy of cortical areas, increasing evidence points towards a reverse process, with higher-order cognitive influences interacting with information coming from the retina.Thus, rather than having a fixed functional role, neurons should be thought of as adaptive processors, changing their function according to the behavioural context.Vision is an active process in which higher-order cognitive influences affect the operations performed by cortical neurons.Visual pathways operate bidirectionally, with each feedforward connection being matched by feedback or re-entrant connections going from higher- to lower-order cortical areas.Top-down influences include various forms of attention, such as spatial, object oriented and feature oriented attention.Top-down influences are not limited to attention but mediate a much broader range of functional roles, including perceptual task, object expectation, scene segmentation, efference copy, working memory and the encoding and recall of learned information.The effect of top-down influences is to change the information conveyed by neurons, both by altering the tuning of their responses to stimulus attributes and by changing the structure of correlations over neuronal ensembles.All areas of the visual pathway, except for the retina, are subject to top-down influences, including early cortical stages of visual processing such as the primary visual cortex and the lateral geniculate nucleus, and all areas along the dorsal and ventral visual cortical pathways. Each area contains an association field of potential interactions, and expresses a subset of these interactions to execute different functions.The sources of top-down influences are widespread, with each area providing information reflecting the functional properties of that area. As a consequence, even a single neuron can be viewed as a microcosm of activity occurring throughout the visual pathway.We propose that the circuit mechanism of top-down control and adaptive processing involves a gating of intrinsic cortical circuits within an area mediated by long-range feedback connections to that area. By selecting a subset of inputs, a neuron can express different components of its association field, and as a result take on different functional roles.}, number={5}, journal={Nature Reviews Neuroscience}, author={Gilbert, Charles D. and Li, Wu}, year={2013}, month=may, pages={350–363}, language={en} }

@article{Jordan_Keller_2020, title={Opposing Influence of Top-down and Bottom-up Input on Excitatory Layer 2/3 Neurons in Mouse Primary Visual Cortex}, volume={108}, ISSN={08966273}, DOI={10.1016/j.neuron.2020.09.024}, abstractNote={Processing in cortical circuits is driven by combinations of cortical and subcortical inputs. These inputs are often conceptually categorized as bottom-up, conveying sensory information, and top-down, conveying contextual information. Using intracellular recordings in mouse primary visual cortex, we measured neuronal responses to visual input, locomotion, and visuomotor mismatches. We show that layer 2/3 (L2/3) neurons compute a difference between top-down motor-related input and bottom-up visual ﬂow input. Most L2/3 neurons responded to visuomotor mismatch with either hyperpolarization or depolarization, and the size of this response was correlated with distinct physiological properties. Consistent with a subtraction of bottom-up and top-down input, visual and motor-related inputs had opposing inﬂuence on L2/3 neurons. In infragranular neurons, we found no evidence of a difference computation and responses were consistent with positive integration of visuomotor inputs. Our results provide evidence that L2/3 functions as a bidirectional comparator of top-down and bottom-up input.}, number={6}, journal={Neuron}, author={Jordan, Rebecca and Keller, Georg B.}, year={2020}, month=dec, pages={1194-1206.e5}, language={en} }

@article{lillicrap_random_2016,
	title = {Random synaptic feedback weights support error backpropagation for deep learning},
	volume = {7},
	copyright = {2016 The Author(s)},
	issn = {2041-1723},
	url = {https://www.nature.com/articles/ncomms13276},
	doi = {10.1038/ncomms13276},
	language = {en},
	number = {1},
	urldate = {2023-09-29},
	journal = {Nature Communications},
	author = {Lillicrap, Timothy P. and Cownden, Daniel and Tweed, Douglas B. and Akerman, Colin J.},
	month = nov,
	year = {2016},
	note = {Number: 1
Publisher: Nature Publishing Group},
	pages = {13276},
}

@misc{portes_distinguishing_2022,
	title = {Distinguishing {Learning} {Rules} with {Brain} {Machine} {Interfaces}},
	url = {http://arxiv.org/abs/2206.13448},
	doi = {10.48550/arXiv.2206.13448},
	urldate = {2023-09-29},
	publisher = {arXiv},
	author = {Portes, Jacob P. and Schmid, Christian and Murray, James M.},
	month = oct,
	year = {2022},
	note = {arXiv:2206.13448 [cs]},
}

@misc{ororbia_learning_2023,
	title = {Learning {Spiking} {Neural} {Systems} with the {Event}-{Driven} {Forward}-{Forward} {Process}},
	url = {http://arxiv.org/abs/2303.18187},
	doi = {10.48550/arXiv.2303.18187},
	urldate = {2023-09-29},
	publisher = {arXiv},
	author = {Ororbia, Alexander},
	month = mar,
	year = {2023},
	note = {arXiv:2303.18187 [cs]},
}

@misc{ororbia_predictive_2023,
	title = {The {Predictive} {Forward}-{Forward} {Algorithm}},
	url = {http://arxiv.org/abs/2301.01452},
	doi = {10.48550/arXiv.2301.01452},
	urldate = {2023-09-29},
	publisher = {arXiv},
	author = {Ororbia, Alexander and Mali, Ankur},
	month = apr,
	year = {2023},
	note = {arXiv:2301.01452 [cs]},
}

@incollection{Jiang2022,
  doi = {10.1093/acrefore/9780190264086.013.328},
  url = {https://doi.org/10.1093/acrefore/9780190264086.013.328},
  year = {2022},
  month = nov,
  publisher = {Oxford University Press},
  author = {Linxing Preston Jiang and Rajesh P.N. Rao},
  title = {Predictive Coding Theories of Cortical Function}
}

@article{bellec_solution_2020,
	title = {A solution to the learning dilemma for recurrent networks of spiking neurons},
	volume = {11},
	copyright = {2020 The Author(s)},
	issn = {2041-1723},
	url = {https://www.nature.com/articles/s41467-020-17236-y},
	doi = {10.1038/s41467-020-17236-y},
	language = {en},
	number = {1},
	urldate = {2023-11-22},
	journal = {Nature Communications},
	author = {Bellec, Guillaume and Scherr, Franz and Subramoney, Anand and Hajek, Elias and Salaj, Darjan and Legenstein, Robert and Maass, Wolfgang},
	month = jul,
	year = {2020},
	note = {Number: 1
Publisher: Nature Publishing Group},
	pages = {3625},
}

@article{murray_local_2019,
	title = {Local online learning in recurrent networks with random feedback},
	volume = {8},
	issn = {2050-084X},
	url = {https://doi.org/10.7554/eLife.43299},
	doi = {10.7554/eLife.43299},
	urldate = {2023-11-22},
	journal = {eLife},
	author = {Murray, James M},
	editor = {Latham, Peter and Frank, Michael J and DePasquale, Brian},
	month = may,
	year = {2019},
	note = {Publisher: eLife Sciences Publications, Ltd},
	pages = {e43299},
}

@misc{millidge_predictive_2022,
	title = {Predictive {Coding}: {Towards} a {Future} of {Deep} {Learning} beyond {Backpropagation}?},
	shorttitle = {Predictive {Coding}},
	url = {https://arxiv.org/abs/2202.09467v1},
	language = {en},
	urldate = {2024-05-22},
	journal = {arXiv.org},
	author = {Millidge, Beren and Salvatori, Tommaso and Song, Yuhang and Bogacz, Rafal and Lukasiewicz, Thomas},
	month = feb,
	year = {2022},
}

@article{halvagal_combination_2023,
	title = {The combination of {Hebbian} and predictive plasticity learns invariant object representations in deep sensory networks},
	volume = {26},
	copyright = {2023 The Author(s)},
	issn = {1546-1726},
	url = {https://www.nature.com/articles/s41593-023-01460-y},
	doi = {10.1038/s41593-023-01460-y},
	language = {en},
	number = {11},
	urldate = {2024-05-22},
	journal = {Nature Neuroscience},
	author = {Halvagal, Manu Srinath and Zenke, Friedemann},
	month = nov,
	year = {2023},
	note = {Publisher: Nature Publishing Group},
	pages = {1906--1915},
}

@misc{oord2019representationlearningcontrastivepredictive,
      title={Representation Learning with Contrastive Predictive Coding}, 
      author={Aaron van den Oord and Yazhe Li and Oriol Vinyals},
      year={2019},
      eprint={1807.03748},
      archivePrefix={arXiv},
      primaryClass={cs.LG},
      url={https://arxiv.org/abs/1807.03748}, 
}

@misc{chen2020simpleframeworkcontrastivelearning,
      title={A Simple Framework for Contrastive Learning of Visual Representations}, 
      author={Ting Chen and Simon Kornblith and Mohammad Norouzi and Geoffrey Hinton},
      year={2020},
      eprint={2002.05709},
      archivePrefix={arXiv},
      primaryClass={cs.LG},
      url={https://arxiv.org/abs/2002.05709}, 
}

@misc{he2020momentumcontrastunsupervisedvisual,
      title={Momentum Contrast for Unsupervised Visual Representation Learning}, 
      author={Kaiming He and Haoqi Fan and Yuxin Wu and Saining Xie and Ross Girshick},
      year={2020},
      eprint={1911.05722},
      archivePrefix={arXiv},
      primaryClass={cs.CV},
      url={https://arxiv.org/abs/1911.05722}, 
}

@misc{grill2020bootstraplatentnewapproach,
      title={Bootstrap your own latent: A new approach to self-supervised Learning}, 
      author={Jean-Bastien Grill and Florian Strub and Florent Altché and Corentin Tallec and Pierre H. Richemond and Elena Buchatskaya and Carl Doersch and Bernardo Avila Pires and Zhaohan Daniel Guo and Mohammad Gheshlaghi Azar and Bilal Piot and Koray Kavukcuoglu and Rémi Munos and Michal Valko},
      year={2020},
      eprint={2006.07733},
      archivePrefix={arXiv},
      primaryClass={cs.LG},
      url={https://arxiv.org/abs/2006.07733}, 
}

@article{picciotto2012acetylcholine,
  title={Acetylcholine as a neuromodulator: cholinergic signaling shapes nervous system function and behavior},
  author={Picciotto, Marina R and Higley, Michael J and Mineur, Yann S},
  journal={Neuron},
  volume={76},
  number={1},
  pages={116--129},
  year={2012},
  publisher={Elsevier}
}

@article{vogels2011inhibitory,
  title={Inhibitory plasticity balances excitation and inhibition in sensory pathways and memory networks},
  author={Vogels, T.P. and Sprekeler, H and Zenke, F and Clopath, C and Gerstner, W},
  journal={Science},
  volume={334},
  number={6062},
  pages={1569--1573},
  year={2011},
  publisher={American Association for the Advancement of Science},
  doi={10.1126/science.1211095}
}

@article{turrigiano2004homeostatic,
  title={Homeostatic plasticity in the developing nervous system},
  author={Turrigiano, Gina G and Nelson, Sacha B},
  journal={Nature Reviews Neuroscience},
  volume={5},
  number={2},
  pages={97--107},
  year={2004},
  publisher={Nature Publishing Group},
  doi={10.1038/nrn1327}
}

@article{hua2004neural,
  title={Neural activity and the dynamics of central nervous system development},
  author={Hua, Jackie Yuanyuan and Smith, Stephen J},
  journal={Nature Neuroscience},
  volume={7},
  number={4},
  pages={327--332},
  year={2004},
  publisher={Nature Publishing Group},
  doi={10.1038/nn1218}
}

\newpage

\appendix

\end{document}